\documentstyle[aas2pp4]{article}

\def\rhoc{\rho_{\rm c}}
\def\rhot{\rho_{\rm t}}
\def\tdif{\tau_{\rm dif}}
\def\tnl{\tau_{\rm NL}}
\def\VS{V\'azquez-Semadeni}
\def\VSE{V\'azquez-Semadeni, E.}
\def\EVS{V\'azquez-Semadeni, E.}
\def\u{{\bf u}}
\def\Gammad{\Gamma_d}
\def\Gammas{\Gamma_s}

\received{15 December 1995}
\accepted{July 1996}
\journalid{337}{15 January 1989}
\articleid{11}{14}
\slugcomment{Submitted to {\it The Astrophysical Journal}}

\begin{document}

\title{A Search for Larson-Type Relations in Numerical Simulations of the ISM.
Evidence for Non-Constant Column Densities}

\author{Enrique V\'azquez-Semadeni, Javier
Ballesteros-Paredes and Luis F. Rodr\'\i guez }
\affil{e-mail: {\tt enro,javier,luisfr@astroscu.unam.mx}}
\affil{Instituto de Astronom\'\i a, UNAM, Apdo. Postal 70-264, M\'exico, D.\
F.\ 04510, M\'exico}

\begin{abstract}
We present results from a statistical study of clouds in two-dimensional
numerical simulations of the interstellar medium. Clouds are defined
as connected sets of pixels above an arbitrary density threshold. Similarly to
real interstellar clouds, the clouds in the simulations
exhibit a differential mass spectrum $dN(M)/dM \sim M^{-1.44 \pm .1}$ and a
velocity dispersion-size relation $\Delta v \sim R^{0.41 \pm .08}$,
although the latter with
correlation coefficients of only $\sim 0.6$.
%POSTREF
However, the clouds do {\it not} exhibit a clear density-size relation.
At a given mean density, clouds span a range of sizes from the
smallest resolved scales up to a maximum given by a Larson-type relation
$R_{\rm
max} \sim \rho^\alpha$, with $\alpha = -0.81 \pm .15$, although numerical
effects cannot be ruled out as responsible for the latter correlation.
The continuum of
sizes at a given mean density implies a range of column densities
of up to two orders of magnitude.

This result supports the suggestion by
Scalo that the observed relation may be an
artifact of the observational limitations. In this case, the
non-existence of a density-size relation would suggest that the origin
of the $\Delta v$--$R$ relation is not virial equilibrium of clouds that follow
a $\rho \propto R^{-1}$ law. Instead, the $\Delta v$--$R$ relation
can be interpreted as a
direct consequence of a $k^{-2}$ turbulent spectrum, which is
characteristic of a field of shocks, verified in the
simulations. However, we also discuss the possibility that
the clouds are in balance between self-gravity and turbulence, but with a
scatter of at least a factor of 10 in the velocity dispersion-size
relation, and a
scatter of over a factor of 100 in the density-size relation,
according to the equilibrium relation $\Delta v \sim (NR)^{1/2}$. In
this case, the scatter in column density would be larger than the
dynamic range of the simulations (and most observational studies).

%a density-size
%correlation does exist, but with a scatter larger than the column density
%dynamic range of the simulations (and of most observational studies).

Finally, we compare these results with recent observational
data. We propose a simple model suggesting that recent
results in which nearly constant column
densities are derived for
dark IRAS clouds may be an artifact of a
temperature gradient within the clouds induced by external radiative
heating. As a consequence, we emphasize that IRAS surface brightness maps are
not appropriate for measuring column densities.
%END POSTREF

\end{abstract}


\section{Introduction}

Interstellar clouds appear to follow a set
of scaling relations first noticed by Larson (1981),
and then apparently confirmed (with slight modifications)
by a number of other workers. These ``Larson's relations'' have the form
\begin{mathletters}
\begin{eqnarray}
\rho \sim R^\alpha \label{lardens eq}\\
\Delta v \sim R^\beta \label{larvel eq},
\end{eqnarray}
\end{mathletters}
where $R$ is the cloud size, $\rho$ is the gas density, $\Delta v$ is the
velocity dispersion derived from the line widths, and $\alpha$ and $\beta$
are the constant scaling exponents. Additionally, the clouds are
found to exhibit a mass distribution of the form
\begin{equation}
{dN(M)\over dM} \sim M^{n}.\label{masspec eq}
\end{equation}
The most commonly quoted values of the
exponents are $\alpha \sim -1.15 \pm .15 $, $\beta \sim 0.4 \pm .1$,
and $n \sim -1.55 \pm .15$
(\cite{lar81}; \cite{tor83}; \cite{dam86}; \cite{fp87}; \cite{myeg88};
Falgarone, Puget \& P\'erault 1992, hereafter
FPP; \cite{fulm92}; \cite{mieb94}; Wood, Myers
\& Daugherty 1994, hereafter WMD; \cite{casm95};
see also the reviews by Scalo (1985, 1987) and Blitz (1991)). However,
significantly discrepant values have also been reported (e.g., \cite{car87};
\cite{lor89}), and the validity of these scaling relations is
currently the subject of strong controversy within the community.

The above ``standard'' values of the exponents for eqs.\ (\ref{lardens eq})
and (\ref{larvel eq}) have been interpreted in terms of the virial theorem
(e.g., \cite{lar81}; \cite{myeg88}; \cite{casm95}).
For $\alpha=-1$ (which coincidentally implies constant column density),
a value $\beta=0.5$ implies virial balance between self-gravity
and the internal velocity dispersion. However, note that for an arbitrary
value of the density scaling exponent $\alpha$,
a corresponding virial balance value
of $\beta$ can always be found (\cite{vsg95}). Thus, the
density-size relation (\ref{lardens eq}) remains unexplained.
In fact, it has been
proposed by Scalo (1990) that this relation may be a mere artifact of
the dynamic range limitations of the observations, and does not
reflect a real property of interstellar clouds.
%POSTREF
In particular, in the case of molecular line data,
the observations are restricted to column
densities large enough that the tracer molecule is shielded against
photo-dissociating radiation. On the other hand, while the
proportionality between line integrated CO intensity and mass surface
density has been reliably established for extragalactic observations
(\cite{dick86}), this relationship is only valid for scales
at which calibrations have been possible, i.e., scales larger than a
few pc. Furthermore, for
clumps within molecular clouds, the structures identified in CO
often do not correspond to those identified with higher-density
tracers (e.g., \cite{massi94}; J.\ Scalo, private communication).
%END POSTREF

In this paper we present the mass spectrum of clouds and
search for Larson-type correlations
in three two-dimensional numerical
simulations of turbulence in the interstellar medium (ISM), one from
Passot, \VS\ \& Pouquet (1995, hereafter Paper I), and the other two being
variants
of the former with respectively larger density contrasts and larger resolution.
We have developed a cloud-identifying
algorithm which allows us to measure the average density, velocity dispersion
and total mass within well-defined (though arbitrary) cloud boundaries in the
density fields of the simulations. The cloud sample thus obtained
has the advantage over actual observations that no
tracer limitations exist in ``detecting'' the clouds, although, on the
other hand, a form of saturation is present due to the relatively small density
dynamic range and other numerical limitations of the simulations.

In \S\ 2
we briefly describe the simulations and the cloud-identifying algorithm, and in
\S\ 3 we present the resulting statistical cloud properties. In \S\ 4
we analyze the limitations of the simulations,
discuss the implications of the absence of a density-size relation, and give a
preliminary discussion of the physical mechanisms behind the velocity
dispersion-size relation. We also
compare the results with recent corresponding observations, in particular
those of WMD. Finally, \S\ 5 summarizes the results.

\section{Numerical Method}

Paper I presented magneto-hydrodynamical simulations of the ISM incorporating
model terms for cooling, diffuse heating and local heating from star formation.
The simulations solve the equations for
the evolution of the density, velocity, internal energy and magnetic
fields in the presence of self-gravity, namely
%POSTREF
\begin{equation}
{\partial\rho\over\partial t} + {\nabla}\cdot (\rho\u) = \mu
\nabla^2 \rho, \label{contin}
\end{equation}
\begin{eqnarray}
{\partial\u\over\partial t} + \u\cdot\nabla\u =
-{\nabla P\over \rho} - {\nu_8} {\nabla^8\u}
- \Bigl({J \over M_a}\Bigr)^2 \nabla \phi \nonumber \\
+ {1 \over \rho} \bigl(\nabla \times {\bf B}\bigr)
\times {\bf B} - 2 \Omega \times \u, \label{mom}
\end{eqnarray}
\begin{equation}
{\partial e\over\partial t} + \u\cdot\nabla e = -(\gamma -1)
e\nabla\cdot \u + {\kappa}_T {\nabla^2e\over \rho} + \Gammad
+ \Gammas - \rho \Lambda, \label{ener}
\end{equation}
\begin{equation}
{\partial {\bf B}\over\partial t} = \nabla \times (\u \times {\bf B})
- {\nu_8} {\nabla^8{\bf B}}, \label{magn}
\end{equation}
\begin{equation}
\nabla^2 \phi=\rho -1. \label{poisson}
\end{equation}
%END POSTREF
We refer the reader to V\'azquez-Semadeni, Passot
\& Pouquet (1995a)
and Paper I for full details on the simulations. Here we just point out that
the simulations from which the data are extracted represent a square section
of the ISM along the Galactic plane of size 1 kpc on a side, with a resolution
of 512 grid points per dimension.
%POSTREF
Also, it is important to note that all the evolution equations contain
dissipative or diffusive terms which are necessary since the numerical
technique used to solve the equations (spectral method) does not
produce numerical viscosity, so the dissipation must be included
explicitly. The momentum and magnetic field equations contain
``hyperviscosity'' terms of the form $\nabla^8$,
which confine the viscous effects to the
very smallest scales in the simulations (\cite{babetal87};
\cite{McW84}). However, although globally dissipative, hyperviscosity
is not everywhere positive definite (\cite{pp88}),
so standard Laplacian terms are
used in the continuity and internal energy equations.
%END POSTREF

Throughout the paper, densities are expressed in units of 1 cm$^{-3}$ and
velocities in units of 11.7 km s$^{-1}$, the units used in \cite{paperI}.
In particular, in the present paper we will use data from the run
labeled Run 28 in Paper I. However, the
star-formation scheme used in the simulations of Paper I
assumes that a star is formed
wherever the local density exceeds a critical value $\rhoc$. (A
``star'' in the simulations is a point source of heat.)
This naturally imposes an upper limit on the densities
reached by the model, since the stellar heating increases the local
pressure and causes the gas to expand.
Thus, densities above $\rhoc$ are very rarely
reached. In Run 28, $\rhoc = 30$, limiting the density contrast $\rho_{\rm max}
/\rho_{\rm min}$ to values $\sim 1000$. In order to obtain a somewhat
larger dynamic range, we have performed an additional run,
called Run 28bis, which is identical to Run 28 up to $t=6.5 \times
10^7$ yr, but afterwards has the star formation turned off. This run ends
up collapsing gravitationally at $t \sim 8.8 \times 10^7$ yr due to the
lack of support from
stellar energy injection, but intermediate times provide a good
framework for study, exhibiting density maxima $\gtrsim 100$ cm$^{-3}$,
and density contrasts $\gtrsim 5000$.
Finally, a run similar to Run 28bis but at a larger resolution
($800 \times 800$, refered to as Run 28.800) was also performed to discuss the
effects of dissipation.  As an illustration, fig.\
\ref{denfield} shows a contour plot of the density field of Run 28
at $t=6.6 \times 10^7$ yr, with the velocity field represented by the
arrows. Note that, although the
size scales represented by the runs are larger than those
studied by most papers concerned with cloud statistics, we expect the
results to be applicable, since all terms in the equations solved in the
simulations are scale-free, except for the dissipative ones.

In order to investigate the statistical properties of the clouds in the
simulations, we have developed an automated algorithm which identifies and
labels clouds. We define clouds as
connected sets of pixels with densities above a user-defined
density threshold $\rhot$. The types of clouds that are identified in this way
are thus clearly dependent on the value chosen for $\rhot$, a somewhat similar
situation to performing observations using various density tracers.
As an example, figs.\ \ref{mask}a
and \ref{mask}b show the clouds obtained by respectively
setting $\rhot=4$ and $\rhot=16$, at time $t=6.6 \times 10^7$ yr into the
evolution of Run 28.

The extremely irregular shapes of most
clouds are noteworthy, and in fact pose a problem in defining the ``size''
$R$ of a given cloud. For simplicity, here we take the size of a cloud as
the square root of the number of pixels it contains.
This definition
may be somewhat unrealistic if clouds are fractal in such a way that their
perimeters are not proportional to the square root of their areas (e.g.,
\cite{sca90}; \cite{fpw91}), but for simplicity and
similarity to observational work we adopt it here. Also, for convenience,
sizes are expressed in pixels in most of the figures below
(1 pixel $\sim$ 2 pc).

For each cloud found by the algorithm, it is then a trivial matter to
compute the average density, the mass, and the turbulent velocity dispersion,
calculated as $\Delta v \equiv
\langle ({\bf u} - \langle {\bf u} \rangle)^2 \rangle^{1/2}$, where ${\bf u}$
is the local value of the velocity and $\langle {\bf u} \rangle$
is the average over the cloud area. Note that this quantity is not
density-weighted.

\section{Statistical Cloud Properties} \label{stat}

We have obtained relatively large samples of clouds by considering values
of $\rhot=$3, 4, 6, 8, 11, 16, 23, and 29 in Run 28, and $\rhot=$3, 4, 6,
8, 11, 16, 23, 32, 45, 64, and 90 in Run 28bis.
Clouds obtained with each value of
$\rhot$ are indicated with a particular symbol type in the plots
discussed below.
In what follows, the data from Run 28 will be considered at $t= 6.6 \times
10^7$ yr, and Run 28bis at $t=7.15 \times 10^7$ yr, unless otherwise noted. The
cloud samples for those cases contain 158 and 145 clouds, respectively.

Figures \ref{masspec plot}a and \ref{masspec plot}b show the logarithmic
mass distributions of the clouds for Run28 and Run 28bis, respectively.
The masses are expressed in the nondimensional units of the
simulations, in which
the total mass contained is $4 \pi^2$, corresponding to
$0.36 \times 10^7 M_{\sun}$ in real units.
The turnover at low masses may be attributed
to incompleteness due to the finite resolution. The high-mass sides of the
distributions, however, exhibit least-squares slopes (fitted for $\log M > 0.8$
and shown as the straight lines) which
imply indices $n=-1.51$ and $-1.43$, respectively (c.f.\ eq.\
[\ref{masspec eq}]).
These values are within the range of values reported in various observational
results (e.g., \cite{wmd}; see also the reviews by \cite{sca85}, \cite{bli91},
\cite{mun94}, and references therein).

Figures  \ref{larvel plot}a and \ref{larvel plot}b show the
correlation between velocity dispersion and size for all clouds with sizes
$\gtrsim 2$ pixels (1-pixel clouds are excluded as they have no velocity
dispersion). Least squares
fits to the data give slopes (shown as the solid lines)
$\beta \sim 0.37$ and $\beta \sim 0.39$
respectively for Run 28 and Run 28bis,
with moderately tight correlation coefficients $\sim 0.6$. The
large scatter of about one order of magnitude in the
correlations should be emphasized. This is comparable to the scatter
found for molecular-line data (e.g., \cite{fpp}). The derived values of $\beta
\sim 0.4$ are slightly lower than the most commonly accepted value of 0.5
(dotted lines),
but are remarkably close to determinations that include heterogeneous
data sets (\cite{lar81}; \cite{fpp}; \cite{fulm92}).

Figures \ref{lardens plot}a and \ref{lardens plot}b
show the average density $\langle \rho \rangle$
{\it vs.} size for the snapshots of the
two runs. Several points should be remarked. First, note that,
in general, the average density of the clouds is quite similar to the value
of $\rhot$ used to define the cloud, and in fact tends to {\it increase} with
size at each $\rhot$.
This is not surprising, due to the presence of dense ``cores'' inside the
largest clouds at each $\rhot$, which tend to increase their average density.
Second, no clear correlation can be seen in either figure.
Instead, at a given $\rhot$, clouds down to the smallest possible
size are seen. These are {\it small
clouds with low densities} and, therefore, low column
densities\footnote{Note that the column density defined here refers to
  a cut through the clouds {\it on} the plane of the
  simulations, and has units of cm$^{-2}$. To obtain a column density
  with units cm$^{-1}$,
  appropriate to our two-dimensional problem, a multiplication by the
  unit length along the third ($z$) dimension, perpendicular
  to the plane of the simulation, should be performed. For simplicity,
  we omit this constant factor throughout the paper. A similar
  situation applies to the computation of masses.}
$N=\rho R$. Third, the size of the largest cloud at each
$\rhot$ is smaller for larger $\rhot$. In particular, in both figures the set
of
largest clouds at each $\rhot$ seems to lie near a $\rho \sim R^{-1}$ law,
similar
to the standard exponent in Larson's relation (\ref{lardens eq}).

In order to test the robustness of the above results, we performed the same
analysis at various other times in both runs, namely at $t=$7.8, 9.1, 10.4,
11.7 and 13.0$\times 10^7$ yr in Run 28, and $t=$7.8 and 8.6$\times
10^7$ yr in Run 28bis. The average value of the exponent $\beta$ of the
velocity dispersion-size relation is $\langle \beta \rangle = 0.41 \pm .08$,
with
typical correlation coefficients $\sim 0.6$. For the mass spectrum, an average
exponent $\langle n \rangle = -1.44 \pm.1 $ is found. The errors are the
standard deviations of the set of values found
for all times. These results confirm
the fact that the simulations show correlations consistent with the
observations
in both cases.

Regarding the density, in all cases small clouds with low densities exist, the
plots (not shown) being qualitatively similar to figs.\ \ref{lardens plot}a
and \ref{lardens plot}b, and the full ensemble of clouds not exhibiting
any correlation with size. The largest clouds at each $\rhot$, on the
other hand, continue to exhibit a near power-law relation with size, with
average
exponent $\langle \alpha \rangle = -0.81 \pm .16$. This is smaller than the
slopes
found in figs.\ \ref{lardens plot}a and \ref{lardens plot}b, which
coincidentally
seem to have some of the steepest slopes in the distribution. This is
illustrated in fig.\ \ref{large clouds},
which shows density {\it vs.} size for
the largest cloud at each value of $\rhot$  at all
times considered above for the two runs. In this figure, clouds obtained with a
given value of $\rhot$ at any one time in either run are shown with the
same symbol and joined by a dotted line. The resulting curves have been
displaced
by increments of 0.2 in $\log \langle \rho \rangle$ for clarity.
For reference, the solid line shows a $\langle \rho
\rangle \propto R^{-1}$ power law. The three uppermost curves in the figure
correspond to Run 28bis and, because of the
larger density dynamic range of this
run, exhibit power-law behavior over a larger range of scales, while those for
Run 28 saturate at $\rho \sim 30$. For this reason, only clouds with $\log R
>1$
for Run 28bis, and with $\log R > 1.5$ for Run 28
were considered in computing $\langle
\alpha \rangle$. The implications of these results
are discussed in \ref{disc impl lardens}
For convenience, in what follows we drop the brackets
when referring to the average density of clouds.

Finally, it should be pointed out that, although clouds with low
column densities exist in the simulations, most of the mass still resides in
the largest clouds, since the distribution of cloud sizes at a given
mean density appears to be roughly uniform. However, this is possibly
an effect of the absence of supernovae in the simulations, since the
``expanding HII regions'' included do not have enough power to disrupt
the largest gravitationally bound complexes.

\section{Discussion}

\subsection{Applicability and limitations of the results} \label{limit}

The results presented in \S\ \ref{stat} have important implications, provided
that the simulations are indeed representative of ISM dynamics. That this is
likely to be the case, in spite of their two-dimensionality, is
suggested by the fact that the
simulated ISM reproduces both the velocity dispersion-size relation and the
mass spectrum of the clouds (c.f.\ \S\ \ref{stat}),
as well as other physical properties of the ISM,
such as the mean density of giant cloud complexes (\cite{vspp95a}), the
cloud and intercloud magnetic field strengths (\cite{paperI}),
the rate of formation of massive stars (\cite{vspp95b}), etc. However,
one important possible criticism due to the two-dimensionality, is that a
$\rho \sim R^{-1}$ scaling relation in three dimensions might translate into
a $\rho \sim R^0$ relation in two dimensions, just because of the elimination
of one dimension. This could be at the
origin of the near constancy of $\langle \rho \rangle$
observed at every value of $\rhot$.
Closer examination shows that this argument is invalid. Even
though the clouds seem to have nearly constant densities at each value of
$\rhot$, this only reflects the fact that, due to our cloud-identifying
algorithm, small, dense clouds are not
seen at small $\rhot$, since they are ``hidden'' within the larger, low density
ones.
But obviously the density is not constant in the
simulations, invalidating the possibility of an $R^0$ dependence of the
density.

%POSTREF
Another important source of concern is that a significant
fraction of the clouds in the samples have sizes of only a few pixels,
and their properties are thus likely to be affected by viscosity and
diffusion. Thus, it is important to quantify the extent to which the
results of the previous section might be influenced by these terms.
In particular, the question arises as to whether
the existence of low-column density, small clouds might be an artifact
of the dissipative terms.

Concerning viscosity, owing to the hyperviscosity
scheme with a $\nabla^8$ operator, its
effects on the velocity field are confined to wavenumbers
$k$ in the range $k_{\rm max}/2 \lesssim k \leq k_{\rm max}$
with $k_{\rm max}=255$ for Runs 28 and 28bis, as can be
seen from the spectrum in fig.\ \ref{spectra} (see \S \ref{veldisp sect}).
The same applies to the
dissipative term in the magnetic equation (see fig.\
5 of \cite{paperI}). Naively, one would then expect the
effects of viscosity to be present at scales up to twice the smallest
scale of the simulation, i.e., from one to two pixels. Actually, the
correspondence between scale
ranges in real and Fourier spaces is not as sharp, and one can expect
``leakage'' up to possibly 4 pixels. Visual inspection of the
velocity field confirms that shocks are spread over typically 4
pixels. Nevertheless, note that viscosity is not effective if the
velocity gradients are not large, and thus the 4-pixel estimate is an
upper limit to the sizes affected by viscosity.
It is worth pointing out that in Paper I the range of influence of
viscosity was estimated at $\sim 5$ pixels. However, this was an
over-conservative estimate not based on a detailed analysis of shock
widths, and the 4-pixel figure given here should be considered as a slightly
more precise estimate based on the above considerations.

In order to correct for viscosity
effects, clouds with sizes up to 4 pixels in size are excluded from
the ``corrected'' plots shown in the Appendix. Note that, since
clouds have in general elongated shapes and sizes are computed as the
square root of the number of pixels, the possibility exists that
clouds with computed sizes larger than 4 pixels will still be 4 pixels
or less across one particular direction.  However, we believe this
effect may be roughly compensated by the fact that the 4-pixel estimate
is an upper limit, and thus we retain all clouds with sizes larger
than this.

Of greater concern are the possible effects of mass diffusion, since
the standard Laplacian diffusive operator used in the continuity
equation
causes diffusion to be important over a larger range of
scales than the hyperviscosity.
Indeed, the characteristic diffusion time $\tdif$ can be shorter than the
turbulent crossing time $\tnl$ for clouds smaller than about 16 pixels (see
Appendix), and diffusion may dominate over turbulent advection for
those clouds.
In particular, it is possible that the small, low-density clouds
reported in the previous section might be a numerical artifact of the
diffusion, which tends to reduce density peaks and spread the clouds
out, or, conversely, to prevent clouds from reaching higher densities
and smaller sizes than they do in the simulations.

Note, however, that the effect of mass diffusion is
exclusively to damp density gradients originated by the turbulent
velocity field, and so diffusion is incapable of forming clouds by
itself; instead, it only modifies the properties of clouds formed by
turbulence or other processes (gravity or the various instabilities
discussed in Paper I).
We can thus obtain a crude estimate of the size and maximum density
a cloud would have in the absence of diffusion
by integrating the diffusion equation backwards in time for initial
conditions corresponding to the clouds in the simulation, over the
length of a nonlinear time $\tnl$. This correction overestimates
the effects of diffusion, since it neglects the advection term
entirely, and the nonlinear time is computed using the turbulent
velocity associated to the size of the cloud, as given by the
turbulent spectrum in fig.\ \ref{spectra}. However, it is
actually turbulent scales larger than a cloud's size which are most
likely to form it (\cite{elm93}; \cite{vspp95b}). In particular, in
the simulations at least, clouds often form at the interfaces of
expanding shells from previous star formation events, which may have
velocities of several kilometers per second, rather than the
significantly smaller velocities ($\lesssim 1$ km s$^{-1}$) indicated
by the spectrum, which is a globally averaged quantity. Nevertheless, for
robustness, we will use the worst-case correction. The details of
this calculation are given in the Appendix. There it is shown that the
central density of a cloud varies by factors of 3--5 in the worst
cases under the influence of diffusion. Using this ``correction''
factor and assuming the clouds move along lines of constant mass in
the density-size plot, one can produce a ``corrected'' such diagram,
shown in fig.\ \ref{correction} in the Appendix for Run 28bis. There it can be
seen that, although clouds are indeed brought slightly closer to a
correlation, at the lowest average densities cloud sizes still vary by
factors of about 100, maintaining the conclusions from \S \ref{stat}.

As a further test, we have also produced a preliminary
higher-resolution run, labeled
Run 28.800, similar to Run 28bis, but at a resolution of $800 \times
800$ pixels, the largest that we can perform in the CRAY YMP of DGSCA,
UNAM. The density field for this run is shown in fig.\ \ref{dens28.800}.
In order to produce this run, the data from Run 28 at
$t=6.5 \times 10^7$ yr were interpolated to produce initial conditions
for the $800 \times 800$ simulation, and then evolved for another
$0.65 \times 10^7$ yr, enough to develop the additional small scale structure
corresponding to the larger resolution.
Incidentally, it is worth noting that, even
though Run 28.800 has a resolution only $\sim$60\%
larger than Runs 28 and 28bis,
the computational effort it requires is roughly 5 times larger in CPU run
time.
Fig. \ref{lardens800 raw}
shows the resulting density-size raw plot for this
run, including cloud sizes down to 1 pixel.
In the Appendix, fig.\ \ref{correction_800} shows the
corresponding plot incorporating the corrections described above,
namely elimination of clouds with sizes $\leq 4$ pixels, and the
correction for diffusion. In order to maximize the available dynamic
range, in the figures in the Appendix
we have used values of the density threshold as low as
$\rhot=0.5$.
For Run 28.800, it is seen that the clouds with the lowest average
densities span a range of roughly a factor of 200 in sizes after the
corrections, while the raw  data exhibit a range of a factor $\sim 500$.

In summary, the discussions from this section
suggest that the existence of small clouds with low
densities described in \S \ref{stat} is a true consecuence of the
dynamics and not an artifact of the dissipative terms used to
stabilize the equations.
%END POSTREF

\subsection{Implications of the results}

\subsubsection{Density, size and equilibrium} \label{disc impl lardens}

With the above considerations in mind, we then have a number of direct
implications. First, as already stated above, the absence of a density-size
relation implies non-constant column densities for the clouds.
Specifically, cloud column densities vary by over two
orders of magnitude in the simulations. Additionally,
the result that clouds with sizes down to the smallest scales exist at all
values of the mean density implies that the observed density-size relation
may indeed be
%POSTREF2
a product of the limited dynamic range of typical surveys.
%that may not have the sensitivity
%to detect weak lines.
Under these conditions, the derived column
densities could appear to show variation over a modest range only,
while the observed sizes might span a range larger than three orders of
magnitude, thus creating an apparent correlation.
Observational effects as a possible origin of the density-size
relation were
%an artifact of the dynamic range limitations of the observations,
%which tend to select constant-column density clouds,
%END POSTREF2
first suggested by Larson (1981) himself, and later
discussed in more detail by Scalo (1990).
Another reason for the appearance of a spurious density-size
relation may be a selection effect introduced by the tendency of observational
work to focus primarily on global intensity maxima of the maps,
%POSTREF2
therefore possibly missing weaker, local maxima, a bias that only recently
has started to be avoided (e.g., \cite{fpp}).
%END POSTREF2

The Larson-type relation defined by the largest clouds, with a scaling exponent
$\alpha \sim -0.8 \pm .16$ is particularly interesting.
The immediate question that
arises is whether this relation is physical, or is induced by numerical
constraints.
On the physical side, a first consideration is that,
if the clouds are hierarchically nested
(smaller, denser clouds are part of larger, less-dense ones), then mass
conservation implies $-\alpha < 3$ in three dimensions. In our two-dimensional
case, this limit becomes $-\alpha < 2$. This is a physical limit which
in the simulations is pushed closer to the observed relation because of the
two-dimensionality.

Another physical issue is whether the large clouds which follow a density-size
scaling law are virialized, so that the standard scenario in which Larson's
relations hold for virialized clouds would apply to the largest clouds.
However,
examining cloud virialization in our simulations turns out not to be a
straightforward task(\S \ref{veldisp sect}).
%POSTREF2
%Deleted sentence
%END POSTREF2
Here, we just check whether the standard $\Delta v$-$R$ relation is also
satisfied in the large clouds. To this end,
some clouds in figs.\ \ref{larvel plot} and \ref{lardens plot}
have been labeled with numbers
so that they can be identified from one figure to another.
Interestingly, for Run
28 at $t=6.6 \times 10^7$ yr (case {\it a} in the figures), the largest clouds,
which are very close to the constant-column density line $\rho \sim R^{-1}$,
shown as the straight line in fig.\ \ref{lardens plot}a, are also very close
to the line $\Delta v \sim R^{1/2}$ in fig.\ \ref{larvel plot}a,
suggesting balance between
gravity and turbulence for this set of clouds. However, this is not the case
for the clouds in Run 28bis at $t=7.15 \times  10^7$ yr (case b in the
figures). In this case, although the clouds again show a slope very close
to $-1$ (fig.\ \ref{lardens plot}b), it can be seen from fig.\
\ref{larvel plot}b that they all have comparable velocity dispersions.
We conclude that even clouds with a $\rho \propto
R^{-1}$ density dependence are not necessarily in equilibrium between
turbulence
and self-gravity. Possibly, magnetic support is more important for
the latter set of clouds, as in the results of Myers \& Goodman (1988).

%POSTREF
An interesting question is whether the various clouds along the
large column density
``envelope'' of the distribution in the density-size relation are
essentially the same cloud seen at various different thresholds
$\rhot$, or else they are truly different clouds. In fact, the
answer is that they are neither.
%POSTREF2
%Deleted sentence
%END POSTREF2
This is
exemplified in fig.\ \ref{hierstr}, which shows a few selected branches
of the cloud hierarchy for Run 28bis. One branch off the largest cloud
includes all the clouds along the envelope, but both the largest and second
largest clouds are seen to also have other branches to daughter clouds
off the envelope (dotted lines).
The same is true of clouds lying immediately below
the envelope (dashed lines), which seem
to define a second envelope of similar slope (see fig.\ \ref{lardens plot}).
%END POSTREF

On the numerical side, the
mass diffusion term in the continuity equation
may tend to reduce the column density of clouds defined through the
threshold density-criterion we use here, since the diffusion
widens and smooths clouds,
whose outer parts may then be left out of the domain defined by $\rhot$.
This effect, plus the plain limitations imposed by the resolution,
clearly prevent the formation of very small clouds with very large
column densities, causing shallower slopes
of the high-column density envelope. In fact,
while the average slope we obtain implies smaller column densities for
smaller clouds, it has been pointed out by Scalo (1985, sect.\ III.A) that
it is obvious from inspection of Lynds' dark cloud catalog
that smaller clouds are darker.
%POSTERF2
Indeed, in our case, the correction discussed in \ref{disc impl
  lardens} and in the Appendix tends to bring this envelope towards
steeper slopes.
%END POSTREF2
In summary, the
specific slopes defined by the largest clouds in the $\log \rho$--$\log R$ plot
cannot be unambiguously attributed to real physical effects. High-resolution,
three-dimensional simulations are needed to resolve this issue.
%POSTREF
Unfortunately, the largest simulations of supersonic compressible
turbulence known to us
(e.g., \cite{por94}, using $512^3$ grid points), are purely
hydrodynamic and do not contain many essential ingredients of ISM dynamics,
such
as the magnetic field, self-gravity, and stellar (i.e., small-scale,
compressible) forcing. Thus, the necessary calculations are still a few years
in the future.
%END POSTREF

%POSTREF2
In any case, regardless of what the specific slope of the envelope
turns out to be upon removal of numerical effects, our results suggest
that the notion of a density-size scaling relation should probably be
replaced by that of an ``allowed'' region in $\rho$--$R$
space. Whether the high-column density boundary is truly a power law,
the value of its corresponding index, and the physical mechanisms
responsible for it, are issues that remain to be
determined by high resolution 3D simulations.
%END POSTREF2

\subsubsection{Velocity dispersion-size relation and turbulence} \label{veldisp
sect}

Since the density-size relation is not verified for the clouds
in the simulations, yet the dispersion-size relation is, the standard argument
explaining the $\Delta v$--$R$ relation, namely virial
equilibrium in clouds satisfying $\rho \propto R^{-1}$,
cannot be invoked.
%POSTREF
This implication is independent of the dimensionality of the
simulations, as it only relies on the non-existence of a density-size
relation, and not on particular values of the scaling exponents.
%POSTREF2
That is, the two-dimensionality would, at most, change the exponent in the
virial equilibrium relation between $\Delta v$ and $\rho$, but not destroy
the correlation altogether. Since in the simulations no density-size
relation exists, a unique virial equilibrium relation between $\Delta
v$ and $\rho$ does not exist either (although see \S\ \ref{altern}).
%END POSTREF2
Note, however, that a Jeans-type analysis incorporating a ``turbulent
pressure''
(\cite{cha51}; \cite{ssm84}; \cite{bon87}; \cite{elm91}; \cite{vsg95})
such that $\nabla P_{\rm turb} = (\Delta v)^2 \nabla \rho$ gives
\begin{equation}
\Delta v \sim R \rho^{1/2},\label{Jeans eq}
\end{equation}
for clouds of size $R$ equal to their Jeans length. Since the latter is
independent of dimensionality, relation (\ref{Jeans eq}) holds also
independently of dimensionality. Thus, the same scaling laws as in 3D are
expected for two-dimensional clouds
in balance between self-gravity and turbulence.
%POSTREF2
This is consistent with a crude estimate of virial
balance in which one equates the gravitational energy $W$ to the turbulent
kinetic energy $K$. For simple cloud geometries and uniform densities
and turbulent velocity dispersions, in 3D one obtains $GM^2/R \sim M
(\Delta v)^2$. Taking $M \propto \rho R^3$ gives the usual result $(\Delta
v)^2 \propto \rho R^2$. In 2D, on the other hand, $M \propto \rho R^2$,
but the gravitational energy becomes $W \sim G M^2$, thus
preserving the result $(\Delta v)^2 \propto \rho R^2$. However, such a
simplified treatment may not be applicable, since in 2D logarithmic
corrections appear, and also the gravitational
term in the virial theorem
may differ from the gravitational energy. A detailed analysis of
this problem is in progress (\cite{balvs96}). Thus, the instability
analysis is provisionally a more reliable indicator of the equilibrium
relation expected in 2D.
%END POSTREF

In the absence of a density-size relation, a plausible origin of
the velocity dispersion-size relation is the statistical properties of the
turbulence itself. Indeed, an
index $\beta=1/2$ is expected for turbulence
characterized by an energy spectrum $E(k) \propto k^m$ with spectral index
$m=-2$ (e.g., \cite{bon87}; \cite{vsg95}; \cite{pad95}; \cite{fleck96}),
where $k$ is the wavenumber associated
with scale $R=2 \pi/ k$. Such a spectral index is the signature
of a field of random shocks (see \cite{ppw88} and
references therein). Figure \ref{spectra} shows the spectra
for the incompressible (solid line) and irrotational (dashed line) parts of
the velocity fields of fig.\ \ref{denfield}. For comparison,
the straight dotted line shows a $k^{-2}$
power law. It is clearly seen that the spectrum of the incompressible
component is remarkably well described by
this slope. The irrotational, or compressible,
component exhibits somewhat stronger fluctuations (most likely due to the fact
that the ``stars'' in the simulations inject purely compressible energy),
but is still very close on the average.
The 20\% discrepancy with the
index $\beta \sim 0.4$ found in the simulations may be due to the fact that
in the simulations there exists an upper bound to the turbulent velocity
dispersion (of order a few km s$^{-1}$) that can be imparted to the medium by
the stellar heating, since in the model they only heat the gas to $\sim 10^4$
K.
This introduces a ``truncation'' on the $\Delta v$-$R$ relation, which flattens
the resulting slope, as can be seen in figs.\ {\ref{larvel plot}a
and \ref{larvel plot}b.

%POSTREF
\subsubsection{An alternative interpretation} \label{altern}

At this point, one important alternative must be pointed out.
%END POSTREF
The large scatter in the velocity dispersion-size relation
would be consistent with the clouds {\it not} having
constant column densities, even if they were in equilibrium between turbulence
and self-gravity. Indeed, in equilibrium, the scatter of about one
order of magnitude in the velocity dispersion would imply
a scatter of roughly two orders of magnitude in the column density,
as can be seen from the equilibrium relation
\begin{equation}
\Delta v \sim (N R)^{1/2},\label{virial eq}
\end{equation}
which is equivalent to relation (\ref{Jeans eq}). Thus, our results can
also be interpreted in the sense that all clouds tend to be virialized,
although with a scatter
of at least two orders of magnitude in the column density. Preliminary
results on the energy budget of a smaller cloud sample in Run 28
(\cite{balvs95})
suggest that the sum of the kinetic and magnetic energies is within one order
of magnitude of the gravitational energy, although surface terms were not
considered there. Also, the total
gravitational and the turbulent kinetic energies per unit mass in the
simulations are almost in equipartition, as shown in fig.\
\ref{egravkin}

In this alternative interpretation, a
density-size correlation may still be present, but missed by the simulations
because the scatter is larger than the column
density dynamic range in the simulations (and in most observational
studies).  This scenario
would leave the origin of the putative density-size relation unexplained,
although it may still be possible that the $\Delta v$-$R$ relation is
originated
by the turbulent energy spectrum, and that the $\rho$-$R$ relation is the
consequence of virial equilibrium.
Although this scenario cannot be ruled out
with certainty until significantly higher-resolution simulations are
performed, it seems unlikely, because in the simulations, clouds
with sizes down to the smallest resolved scale are often
found even at the lowest values of $\rhot$, thus not giving any indication
of the presence of a density-size relation. Instead, clouds of similar
average densities often span the whole range of scales accessible to
them. Also, the global balance between the turbulent and gravitational
energies is mostly just a consequence of the presence of a
self-regulated cycle of gravitational contraction, star formation,
energy injection to the medium and dispersal, and again gravitational
contraction, as discussed in \VS\ et al.\ (1995a), so that near global balance
between turbulence and self-gravity is maintained at all times.

Finally, note that here we have not discussed other mechanisms that have been
suggested in the literature as responsible for the $\Delta v$-$R$ relation,
such
as inverse cascades of angular momentum, (\cite{hent84}), critical thermal
pressure equilibria (\cite{chie87}) or the contribution
from the magnetic support (\cite{shu87}; \cite{myeg88}; \cite{moups95};
\cite{gam96}).
We will address the role of these processes in the simulations in future work.

\subsection{Comparison with observations}

The lack of correlation between cloud density and size in the simulations
appears
to be in contradiction with the correlations found in most observational
results
(\cite{lar81}; \cite{tor83}; \cite{dam86}; \cite{fp87}; \cite{mye90};
\cite{wmd}).
However, as discussed in \S \ \ref{disc impl lardens}, we interpret
the discrepancy as an effect of the limited column density dynamic range of
the observations or of selection effects introduced
by focusing exclusively on global intensity maxima in the maps.
There are some examples of observations that have intended to avoid these
problems (\cite{car87}; \cite{lor89}; \cite{fpp}).
The first two authors have focused on clumps within a single
molecular cloud with extensive star formation, thus better sampling the
non-gravitationally-bound turbulent transients.
The work of FPP was specifically tailored towards studying
low-bright\-ness regions in molecular clouds. In both Loren (1989) and FPP,
column densities spanning over one and a half orders of magnitude are found.

On the other hand, WMD have recently concluded from an
analysis of 60~$\mu$m and 100~$\mu$m IRAS maps, that column densities
of dark clouds cluster typically at N(H$_2$) $\simeq$ 4 $\times$10$^{21}$
cm$^{-2}$ (corresponding to a typical 100~$\mu$m opacity of
$\tau_{100}$ $\simeq$ 200 $\mu$Nepers), while claiming that the dynamic
range of the data would have allowed detection of any significant variations.
The discrepancy between these results and those of the present paper seemingly
cannot be explained in terms of the limited column density
dynamic range of their observations.

In actuality, we believe that the results of WMD may be spurious and
attributed to a combination of their selection criteria and
the following effect. At 60~$\mu$m and 100~$\mu$m, one is observing
``warm'' dust that could be coming from the ``edges'' of molecular clouds.
If this ``warm'' dust is being heated by the visible photons
from the galactic stellar radiation field,
it is expected that the depth of the ``warm'' region will be of order
of a few Nepers in the visible. This will lead to apparent
constant optical depths when determined from observations of the ``warm'' dust.

%POSTREF
Detailed radiative transfer models have been presented by
\cite{bern92}, whom have already warned against using 100~$\mu$m
surface brightness as a tracer of dust column density. In what
follows, we present
a simpler model that allows us to give first-order estimates of the
scaling of the intensities in the IRAS 60 and 100~$\mu$m bands and the
{\it apparent} dust opacity as a function of the true dust opacity
through the cloud.
%END POSTREF

Similarly to \cite{jardh89}
and WMD, we assume that the ratio of visible to
100~$\mu$m absorption efficiency is (Q$_V$/Q$_{100}$) = 2.0$\times$10$^4$.
It is important to emphasize that the 100~$\mu$m opacity
considered in this ratio is being produced by small dust particles
that become heated to relatively large temperatures and that
are responsible for the observed 100- and 60~$\mu$m emission.
The absorption efficiency ratio that includes all the dust at 100~$\mu$m
is about a factor of 10 smaller (\cite{chikk86}).
Furthermore, in the range of wavelengths between 60 and 100~$\mu$m, the
absorption efficiency scales approximately as $\nu^{-1}$. Using this functional
dependence, we can write a crude approximation for the dust temperature as
$$T_{\rm DUST} =(T_{\rm cr}^5 + T_{\rm RAD}^5~e^{-\tau_{\rm V}})^{1/5},$$
where $T_{\rm cr}$ and $T_{\rm RAD}$ are pseudotemperatures
that parameterize the
heatings due to cosmic rays and radiation, respectively. (Although this
approximation is not valid at all wavelengths, we have checked that using
slightly different functional dependences does not significantly alter our
results.) We assume that
cosmic ray heating is constant for any point in the cloud.
Specifically, $T_{\rm cr}$ = 10 K is the temperature that the dust
has if heated only by cosmic rays, and $T_{\rm RAD}$ is the temperature
that the dust would have if heated only by the stellar radiation field
(at the edge of the cloud, where no absorption is present). Also,
$\tau_{\rm V}$ is the opacity in the visible, that increases as
we get deeper into the cloud.
In this simple model $T_{\rm DUST}~\simeq T_{\rm RAD}$ at the edge of the
cloud and it tends to $T_{\rm cr}$ for the inner regions of the cloud (fig.
\ref{figluis1}).
The intensities at 60~$\mu$m and 100~$\mu$m, I$_{60}$ and I$_{100}$, can
be calculated for clouds
with different values of $\tau_{\rm V}$, as shown in fig.\ \ref{figluis2}.
This calculation is made assuming that $Q_{100}/Q_{60} = 60 \mu/100 \mu$. From
this figure it is evident that for clouds exceeding a few Nepers in $\tau_{\rm
V}$, I$_{60}$ and I$_{100}$ do not continue growing with $\tau_{\rm V}$.
An apparent dust temperature, T$_{60/100}$ can
be derived from I$_{60}$ and I$_{100}$ and the far-infrared opacities
are then obtained.
In fig.\ \ref{figluis3}, we plot the 100~$\mu$m opacity derived in this manner
as a function of the cloud's $\tau_{\rm V}$ for different values of $T_{\rm
rad}$.
The apparent 100~$\mu$m opacity ``saturates'' at values in the
order of 70 to 100~$\mu$Nepers. Considering that along a line of
sight one expects to intersect the front and the back edges
of a cloud, the typical 100~$\mu$m opacities of
$\tau_{100}$ $\simeq 200 \mu$Nepers appear to be explained
as a result of this effect. The few clouds with much larger values of
$\tau_{100}$
(up to $\tau_{100} \sim 10^4$) reported by WMD are regions of strong star
formation
activity, which may heat the clouds from the inside, raising $\tau_{100}$.
The presence of this effect is also consistent
with the limb brightening observed in some clouds in the far-infrared by
\cite{shs89} and WMD,
%POSTREF
and predicted by the models of \cite{bern92}.
%END POSTREF
Note that this effect continues to be applicable even if the clouds are clumpy,
since it should hold at the edges of any density peaks, large or small,
as long as they have a large enough column density.
However, note also that this ``saturation'' effect in the
determination of the apparent $\tau_{100}$ will occur
only if reasonably high temperatures ($T_{\rm DUST} \gtrsim 20$ K)
are present at the cloud's surface.
In any case, for lower dust temperatures the emission at 60 and 100~$\mu$m
is very weak and undetectable in practice.
One consequence of this effect is that in order to
fully sample the dust from dark clouds one requires observations at longer
wavelengths that will trace the predominant cooler dust component. We intend to
verify this effect by comparing column densities obtained with different
indicators in future work.

The effect discussed above may clearly impose an upper bound to the
column densities derived by WMD. Furthermore, concerning their selection
criteria, it should be noticed that
WMD define core, cloud and
cloud complex in terms of ranges of extinction. This
directly selects against identification of low-column density
structures. In fact, in their maps, small, low-extinction clouds are readily
seen, but not classified as such precisely because of their low extinctions.
Thus, the column densities of the cores studied by WMD are bound from above
due to the ``saturation'' effect of the cloud edges, and from below by their
very definition of a core, rendering their derived column density constancy
open to question.

\section{Summary and Conclusions}

In this paper we have searched for Larson-type (1981)
correlations and cloud mass spectrum slopes in the clouds generated in
numerical simulations of the ISM, one from Paper I, a similar
one with a larger density dynamic range, and another with higher resolution.
We define a cloud as a connected
set of pixels in the density field with values larger than an arbitrary
threshold $\rhot$. From the results at various
different times in the two simulations,
we find that the mass spectrum has the form $dN/dM \propto M^{-1.44 \pm .1}$,
and the velocity dispersion is related to the cloud size by $\Delta v \propto
R^{0.41\pm .08}$, where the errors
are the standard deviations in the set of values including all
the various times.
The dispersion-size relation exhibits a scatter of about an order of magnitude,
comparable to the scatter observed in real clouds.

The simulated clouds do not exhibit a density-size
relation, but instead, at all mean densities, clouds of sizes down to the
smallest resolved scales are commonly found. This result implies that
the clouds do not have constant column densities, but instead exhibit a range
of roughly two orders of magnitude, thus providing strong support to the
possibility that the observationally-derived
density-size relation is an artifact of the
%POSTREF2
limited dynamic range employed by observational surveys,
%END POSTREF2
and of the criteria used for selecting the clouds, as first noticed
by Larson (1981), and then strongly argued for by Scalo (1990). Our results
suggest that low-column density clouds do exist in the ISM, but are
systematically missed by most observations. In the simulations, low-column
density clouds are turbulent transients, as in the suggestion
by Magnani et al.\ (1993). Observational work that has
used complete cloud samples
(e.g., \cite{lor89}), or specifically looked at structures
in the low brightness regions of molecular clouds (\cite{fpp}), has indeed
found a reported column density variability of over one and a half orders of
magnitude, and masses well below the virial mass, by factors up to two
orders of magnitude. However, note that in our simulations most of the
mass  resides in the largest clouds at each mean density, since there
are comparable numbers of small and large clouds.

The set of largest clouds at every
threshold $\rhot$ exhibits a density-size relation
$\rho \sim R^\alpha$, with $\alpha = -0.8 \pm .15$. These clouds
appear to be close
to balance between turbulence and self-gravity on occasions, but not
in general. On the other hand, pure mass
conservation, which gives an upper limit $-\alpha < 2$ for the two-dimensional
case, together with numerical limitations of the simulations, cannot be ruled
out as the sole causes responsible for this result.
%POSTREF2
In any case, regardless of what the specific slope of the envelope
turns out to be upon removal of numerical effects, our results suggest
that the notion of a density-size scaling relation should probably be
replaced by that of an ``allowed'' region in $\rho$--$R$
space. Whether the high-column density boundary is truly a power law,
the value of its corresponding index, and the physical mechanism
responsible for it, are issues that remain to be
determined by high resolution 3D simulations.
%END POSTREF2

The result that the $\Delta v$-$R$ relation
is verified in the simulations but the
$\rho$-$R$ relation is not, supports
the interpretation that the former
relation is a direct consequence of the statistical properties of the
turbulence, since a turbulent energy spectrum of the form $E(k) \sim k^{-2}$,
as observed in the simulations, implies an $R^{1/2}$ scaling for the velocity
dispersion. However, we discussed the
possibility that a density-size relation does exist, although
with a scatter larger than the column
density dynamic range spanned by the clouds in
the simulations.
%POSTREF
Such a scatter would be consistent with balance
between turbulence and self-gravity, with a scatter of one order of
magnitude in the velocity dispersion, according to relation
(\ref{virial eq}).
In this case, the origin of the $\rho$-$R$ relation would remain
unknown, and the
turbulent origin of the $\Delta v$-$R$ relation would have not as strong a
support.
This possibility cannot be ruled out without very high resolution
simulations in order to add at least another order of magnitude to the
column density dynamic range.
However, the simulations do not give any indication that this may be
the case, since cloud sizes
down to the smallest resolved scales
are found even at the lowest values of $\rhot$. Besides, the column
density dynamic range in the simulations is already larger than that
of most observational surveys.
%END POSTREF

We also discussed the recent results of WMD, who have argued
in favor of a surprisingly constant column density (to within a factor of a
few)
in a large sample of cores studied through IRAS 60 and
100~$\mu$m maps, while claiming a very large dynamic range in their
observational method.
We argued that this result may be spurious, by
presenting a model in which the
warm ``skin'' of clouds and cores, which is likely to always have visual
extinctions of order unity, puts an upper bound to the column densities
measured by WMD. From below, WMD's own selection criteria eliminate
low-column density clouds, thus mimicking a nearly uniform column density.

Finally, we remark that in this paper we have limited the discussion to the
trends of the density and velocity dispersion with size, omitting any
discussion
of the role of the magnetic field, which is clearly important for the dynamics
of the simulations (\cite{paperI}). A detailed account of the energy
budget in the simulations including the magnetic energy density,
as well as surface terms is under way (\cite{balvs96}).

\acknowledgments

We gratefully acknowledge fruitful discussions with A.\ Raga, J.\ Cant\'o,
S.\ Lizano and T.\ Passot, as well as helpful comments and criticisms
from J.\ Scalo. An anonymous referee emphasized the importance of the
effects of dissipation and diffusion.  Runs 28bis and 28.800 were
performed on the Cray YMP 4/64 of DGSCA, UNAM. This work has
received partial financial support from UNAM/CRAY grant SC-002395 and
UNAM/DGAPA grant IN105295.
J. B.-P. acknowledges financial support from a UNAM DGAPA fellowship.

%\appendix
\bigskip
\section{APPENDIX}

In this appendix we make an order of magnitude estimate of the effect of the
numerical
difussion term in the continuity equation.
Consider the continuity equation, eq.\ (\ref{contin}):
\begin{equation}
{\partial \rho \over \partial t} + \nabla \cdot (\rho {\bf u} )= \mu \nabla^2
\rho. \label{mass}
\end{equation}
In order to estimate the effects of difussion,
we consider the extreme case when the
advection term can be neglected, so that the mass equation becomes the
standard difussion equation:
\begin{equation}
{\partial \rho \over \partial t} = \mu \nabla^2 \rho. \label{diff}
\end{equation}
For simplicity, we consider a one-dimensional (axisymmetric) problem,
whose solution is (see, e.g., \cite{hab87}):
\begin{equation}
%\begin{eqnarray}
\rho(x,t) = \sqrt{1 \over 4\pi\mu t} \int_{-\infty}^{+\infty}
f(x') \exp{\biggr(-{(x-x')^2
\over 4 \mu t}\biggl)} dx',
\label{generalsolution}
%\end{eqnarray}
\end{equation}
where $f(x')$ is the initial density distribution, which we assume to
be a Gaussian,
$f(x') = \rho _0 \ \exp(-x'^2 /2\sigma^2)$. In what follows,
we will identify the width $\sigma$ of
the Gaussian with the size of the cloud of interest.
Equation (\ref{generalsolution}) then becomes:
\begin{equation}
\rho(x,t) = \rho_0 \biggr( {1\over t/t_0 +1} \biggl)^{1/2} \exp
\biggr(-{x^2/2\sigma^2
\over t/t_0 +1} \biggl),
\label{particularsolution}
\end{equation}
where
\begin{equation}
\biggr({ t_0 \over t_{\rm code} } \biggl) =
{\sigma^2 \over 2\mu } = 9.41 \times
10^{-3}\biggr({ \sigma \over \rm pixels }\biggl)^2. \label{taudiff}
\end{equation}
is the characteristic diffusion time in
units of the code, $t_{\rm code}=1.3 \times 10^7$
yr, and the second equality follows from using
the value $\mu = 0.008$ (see Paper I) (note that the size of the
integration box is $2\pi$ in the code units).
This value of $t_0$ can be made independent of
the resolution of the simulation by choosing
$\mu$ such that $\mu k_{\rm max}^2={\rm cst,}$ where
$k_{\rm max}$ is the maximum Fourier wave number in the
simulation, equal to 1/2 of the resolution.

Now consider the
turbulent crossing (or nonlinear) time
for scale $\sigma$:
$$\tau_{\rm NL} \sim \sigma / u_l, $$
where $u_l$ is the turbulent speed associated to the scale $l$.
Note that we are allowing
for the possibility that clouds of size $\sigma$
are generated by turbulent streams of
different, typically larger, size $l$ (see \S \ref{limit}).
This velocity can be
estimated from the turbulent energy spectrum $E(k)$ as:
$${1\over 2} u_l^2 = \int_{2\pi/l}^{\infty} E(k) \ dk,$$
that is, $u_l$ is the root mean square energy
per unit mass in scales smaller than $l$.
Using $E(k) = 0.1 k^{-2}$, as indicated by fig.\ \ref{spectra}, one obtains:
\begin{equation}
\biggr( {\tau_{\rm NL} \over t_{\rm code} }\biggl) = 14.05 \biggr( {l/{\rm
pixels}\over
n_{\rm res}/\rm pixels} \biggl)^{1/2},  \label{taunl}
\end{equation}
where $n_{\rm res}$ is the
number of pixels (i.e., the resolution)
per spatial dimension of the simulation. From equations
(\ref{taudiff}) and (\ref{taunl}), we can now
compare the diffusion and nonlinear times, in order to determine
the scale at which they are equal. Assuming that $l=m\sigma$, we obtain:
\begin{equation}
\biggr({l_{\rm eq} \over \rm pixels}\biggl) = {16.33 \over m^{1/2}},
\label{limite}
\end{equation}
and thus, clouds with sizes lower than $\sim 16$
pixels are dominated by the numerical
mass-difussion. Therefore, it is important to
assess the effect of mass diffusion in
the results of \S \ref{stat}. To do this, we
take the extreme position that for time
durations $\Delta t \leq \tau_{\rm NL}$, a
Gaussian cloud is affected exclusively by
diffusion, and compute the fractional variation
of its central density over $\tau_{\rm NL}$.
In fig. \ref{denstime} we show the evolution of
$\rho (0,t)/\rho_0$ given by equation
(\ref{particularsolution}) over $\tau_{\rm  NL}$
for a cloud of size $\sigma =4$ pixels.
Note that this is a worst-case estimate, since
on the one hand, in the plots below we
have discarded clouds with sizes $\leq 4$ pixels
in order to avoid viscosity effects
(see \S \ref{limit}), which is most affected
by diffusion, while, on the other hand, we
have taken $m=1$.

As we can see from fig.\ \ref{denstime},
the density decreases by a factor $\lesssim 5$.
Typically, thus, we can expect our density
data to change by factors smaller than
half an order of magnitude.

Using the above results, we can produce
a ``corrected" density-size plot with the
estimated ``true" densities $\rho_{\rm corr}$ calculated as:
\begin{equation}
\rho_{\rm corr} \sim \rho_{\rm data} ( 1 + \tau_{\rm NL}/t_0 )^{1/2}.
\label{density_correction}
\end{equation}
where $\rho_{\rm data}$ are the raw
cloud mean densities as produced by the simulations,
and the ``true" sizes are obtained assuming
mass conservation:
\begin{equation}
R_{\rm corr}\sim {R_{\rm data} \over (1 + \tau_{\rm NL}/t_0)^{1/4}}.
\label{size_correction}
\end{equation}
The ``corrected" density-size plots are shown in figs.\
\ref{correction} and \ref{correction_800}
for Run 28bis and the high-resolution
Run 28.800, respectively. We see that, although
slightly pushed closer to a correlation
in the case of Run 28 bis, the general trend of
these plots still supports the main
conclusion from \S \ref{stat}, namely that clouds
of sizes down to the resolution are
seen at all mean densities.

\clearpage

\begin{figure}
\plotone{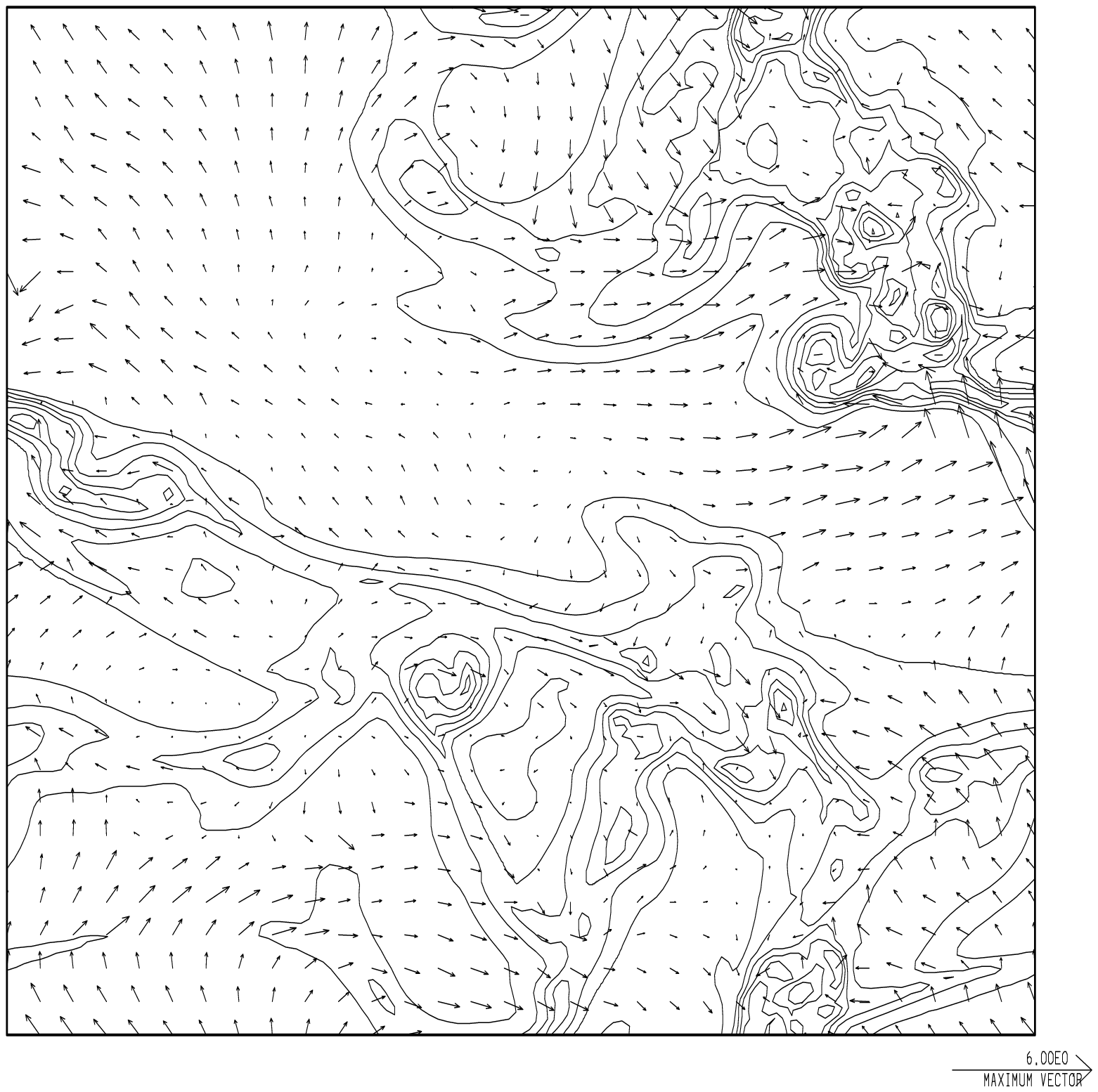}
\vskip 1cm
\caption{Density (contours) and velocity (arrows) fields for Run 28 at
$t=6.6 \times 10^7$ yr. The contours are spaced logarithmically, with
increments
of 0.3125 in $\log \rho$. The minimum density in this plot is $0.04$ cm$^{-3}$,
and the maximum density is 40 cm$^{-3}$.
\label{denfield}}
\end{figure}

\begin{figure}
\plottwo{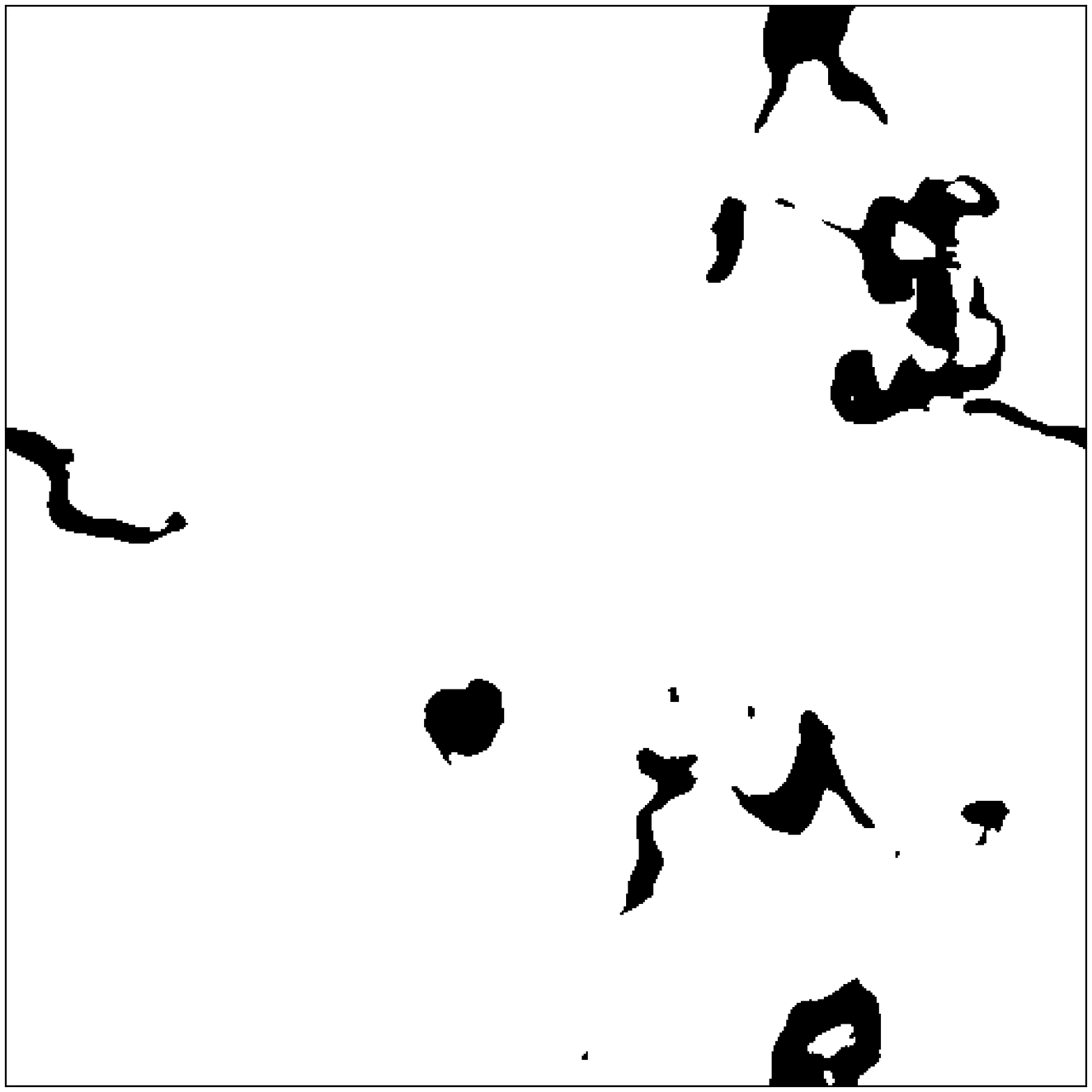}{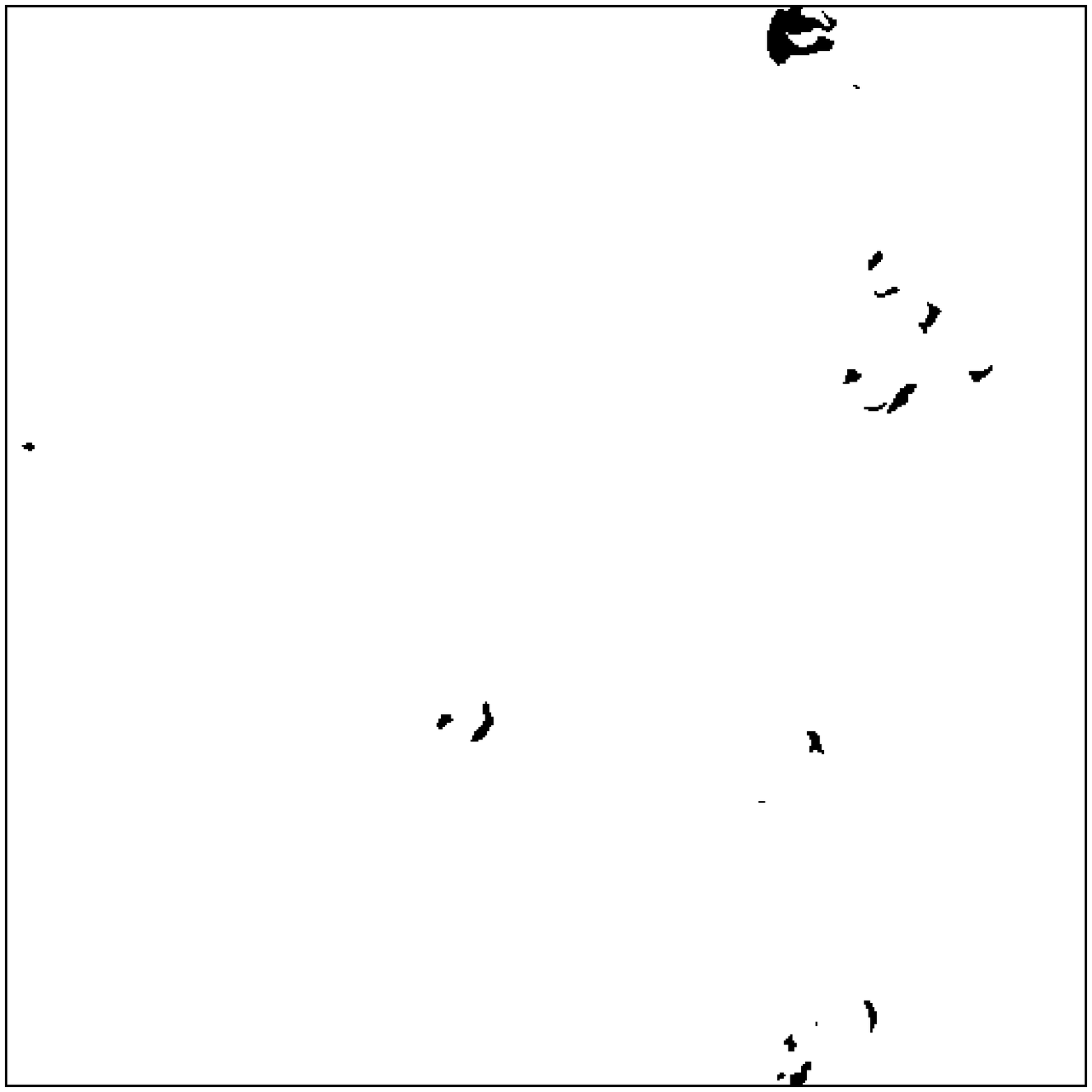}
\caption{Regions (``clouds'') with densities larger than an arbitrary threshold
$\rhot$ in the density field of Run 28 at $t=6.6 \times 10^7$ yr. a) $\rhot=4$.
b) $\rhot=16$. Note that small clouds with high densities
are nested within larger
complexes, but small clouds with low densities can be seen in a) as well.
\label{mask}}
\end{figure}

\begin{figure}
\plottwo{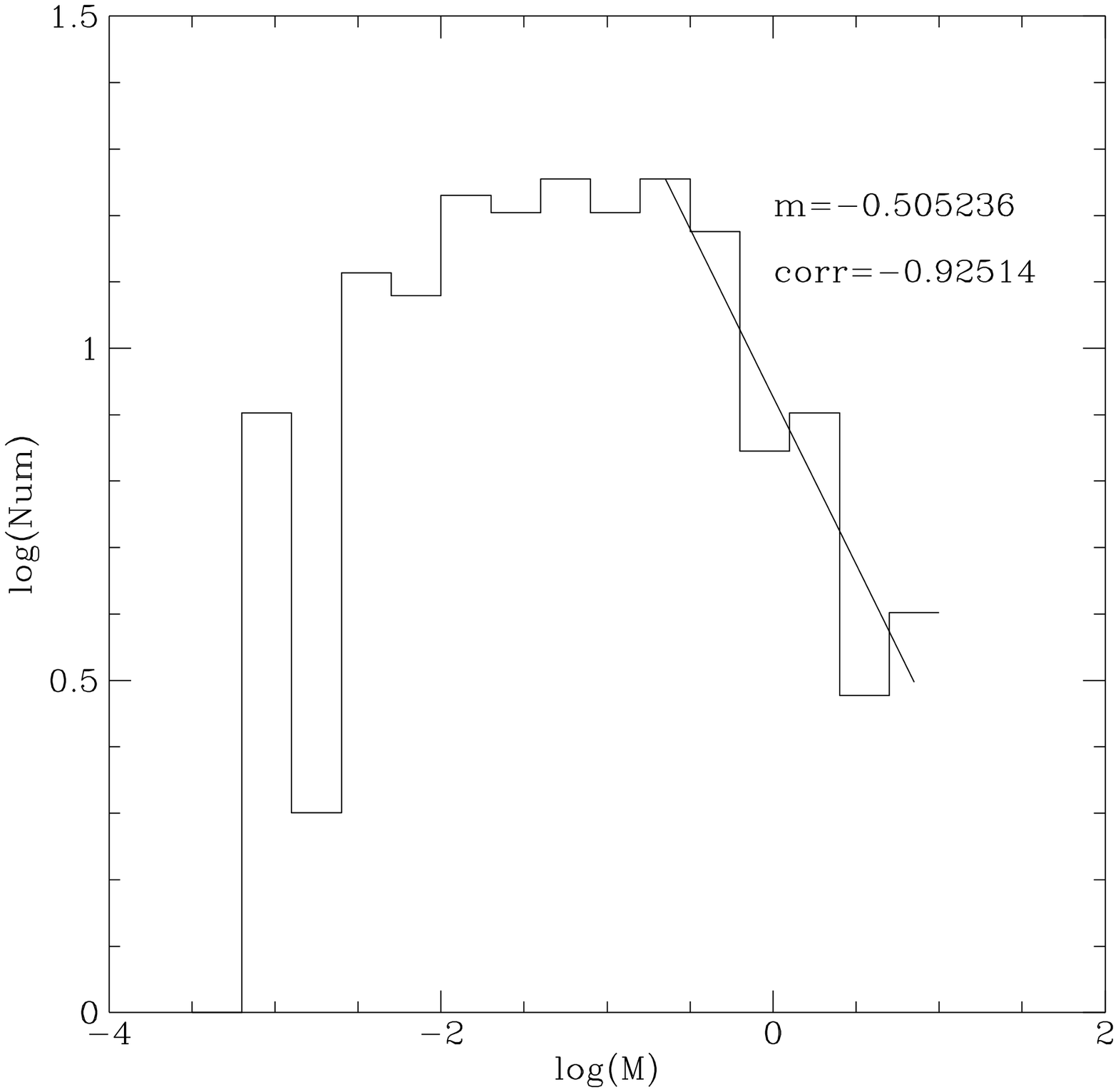}{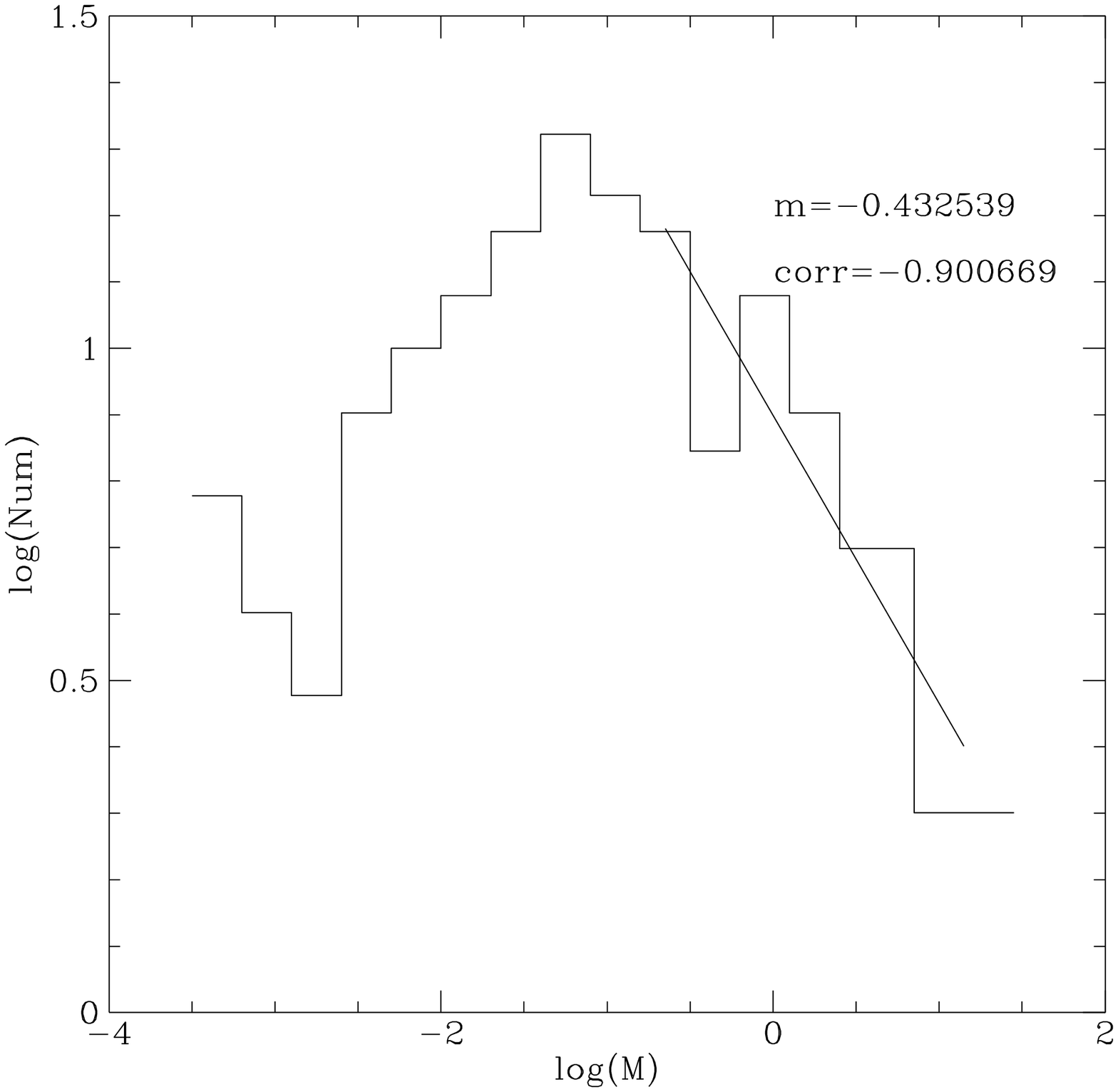}
\caption{Logarithmic mass distributions for a) Run 28 at $t=6.6 \times 10^7$ yr
and b) Run 28bis at $t=7.15 \times 10^7$ yr. The lines show least-squares fits
to data bins with $\log M > 0.8$, and have slopes $-0.51$ in a) and $-0.43$ in
b). \label{masspec plot}}
\end{figure}

\begin{figure}
\plottwo{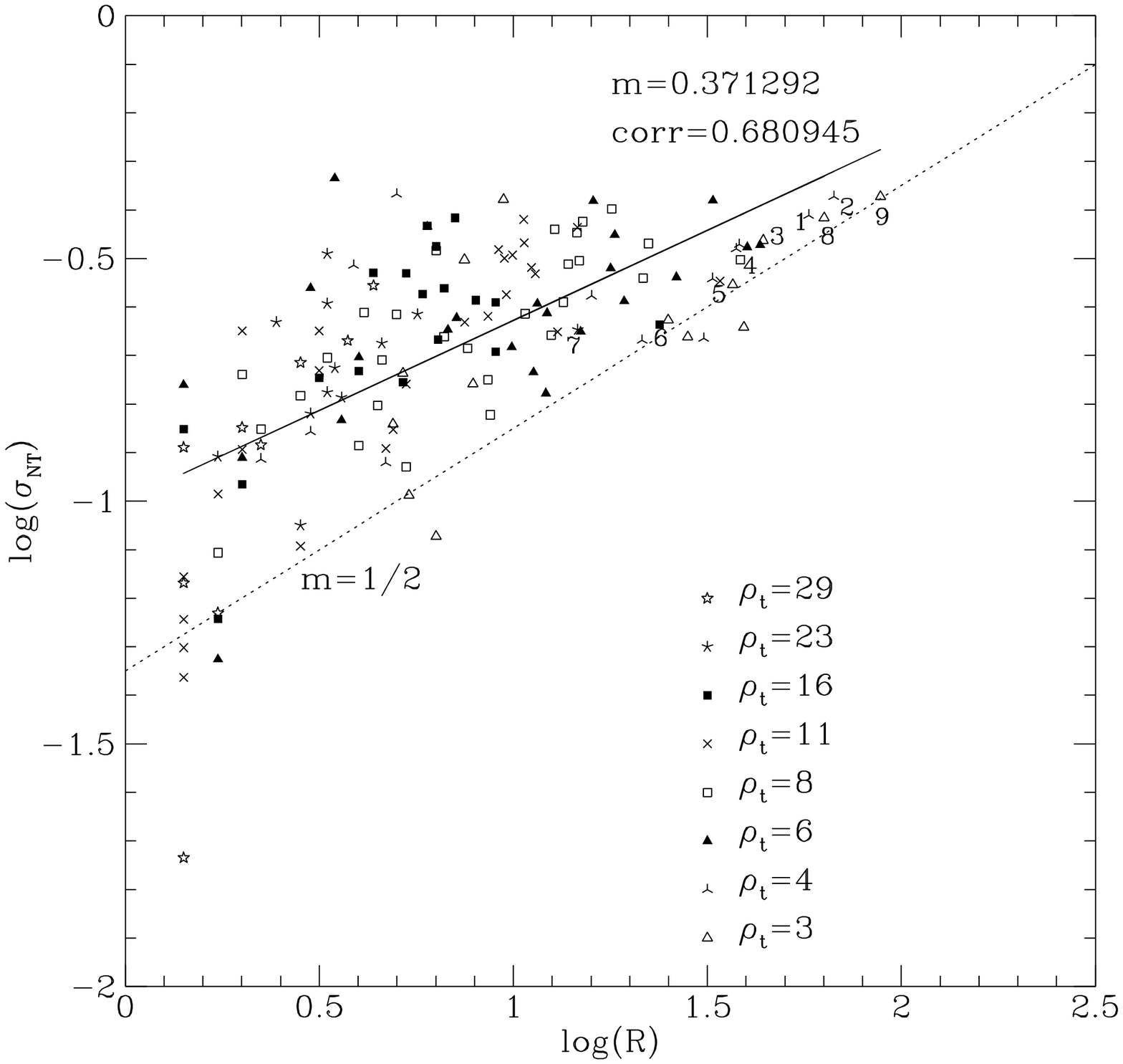}{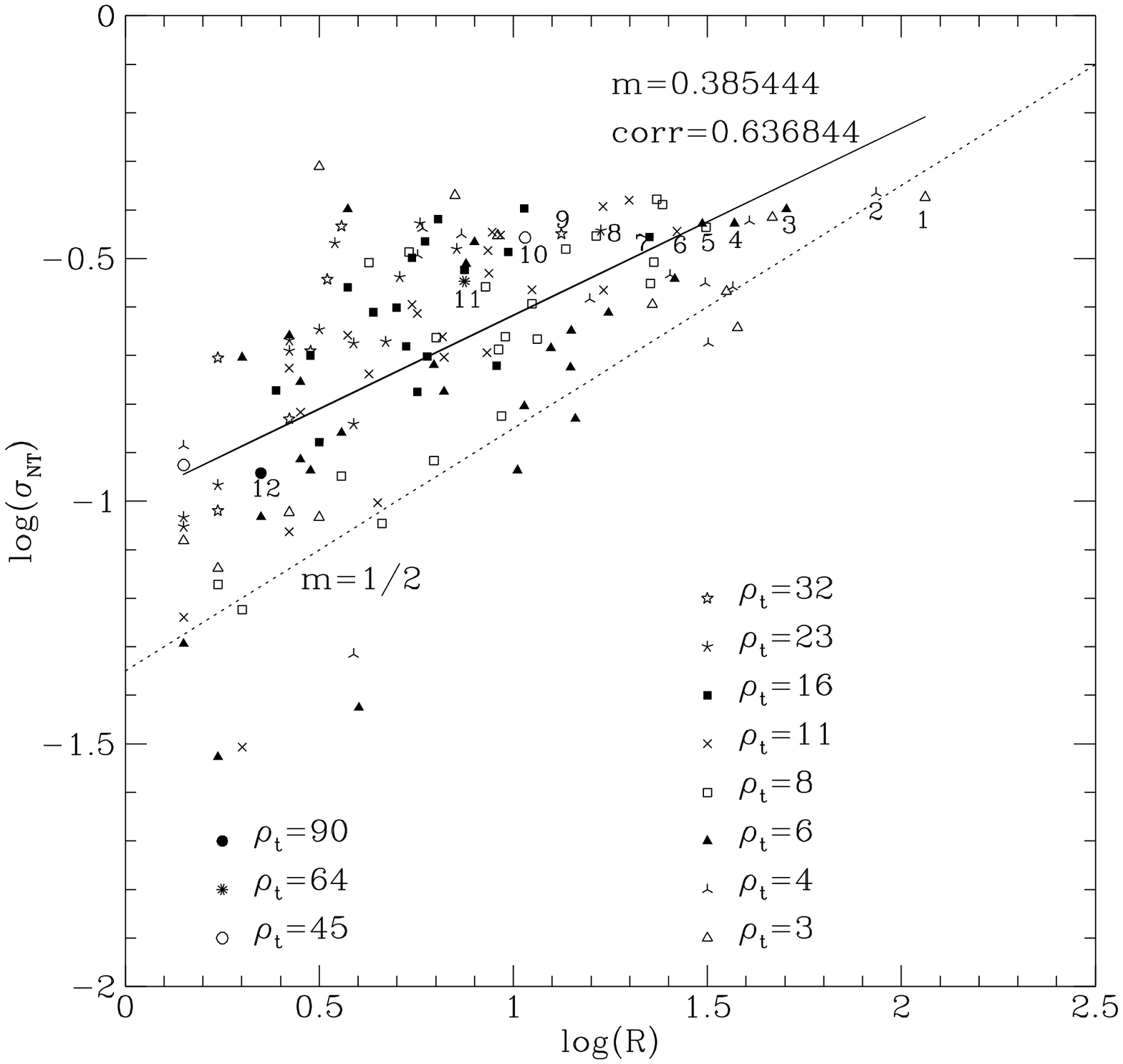}
\caption{Velocity dispersion {\it vs.} size for all clouds in
a) Run 28 at $t=6.6 \times
10^7$ yr and b) Run 28bis at $t=7.15 \times 10^7$ yr. Clouds obtained
with each value of $\rhot$ are shown with a specific symbol as
indicated. The solid lines show
least-squares fits, with the slopes and correlation coefficients indicated.
For reference, a 1/2 slope is indicated by the dotted lines.
The cloud labels in a) and b)
are respectively the same as in a) and b) of figs.\ \protect
\ref{lardens plot}a and b.
\label{larvel plot}}
\end{figure}

\begin{figure}
\plottwo{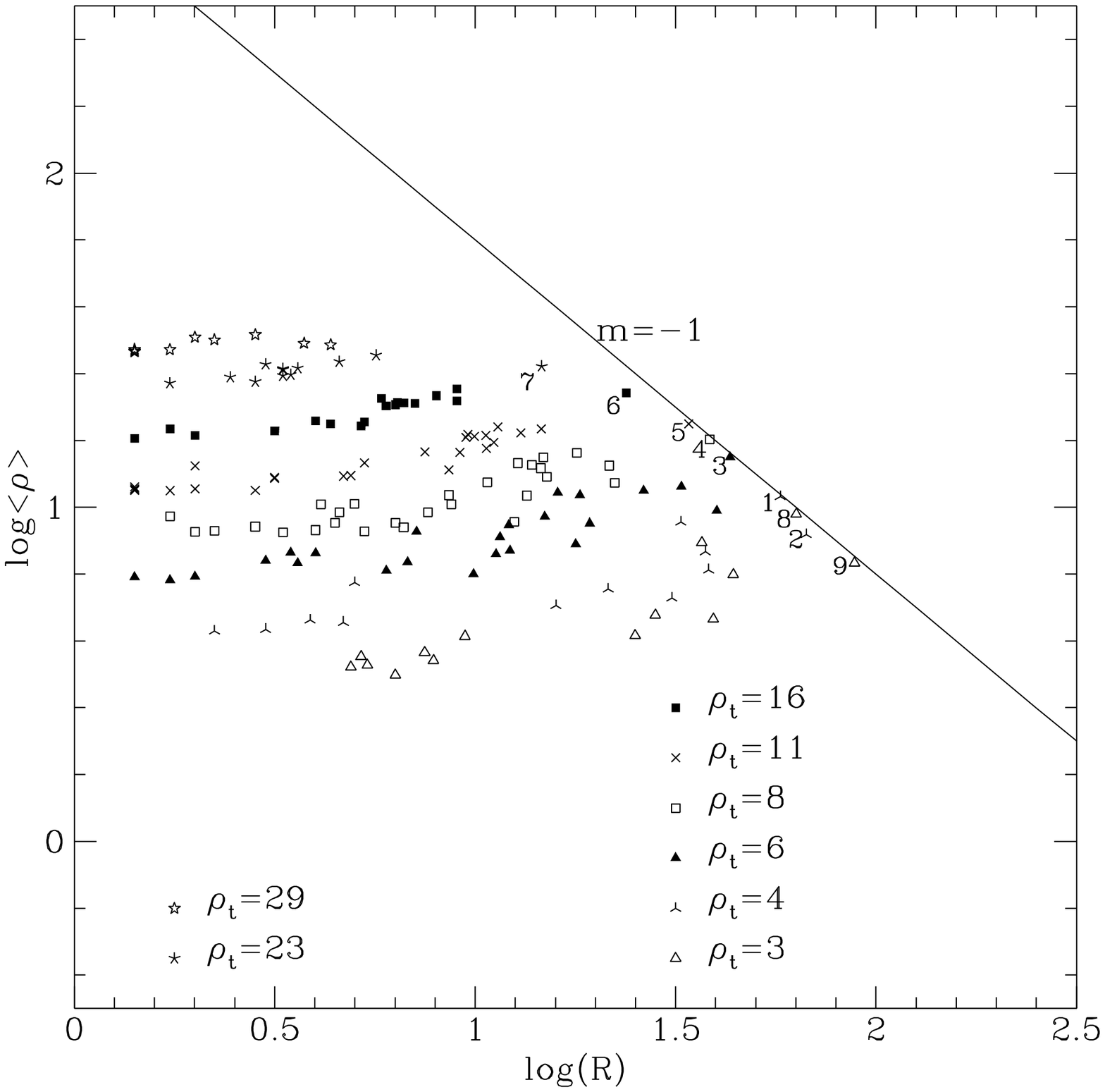}{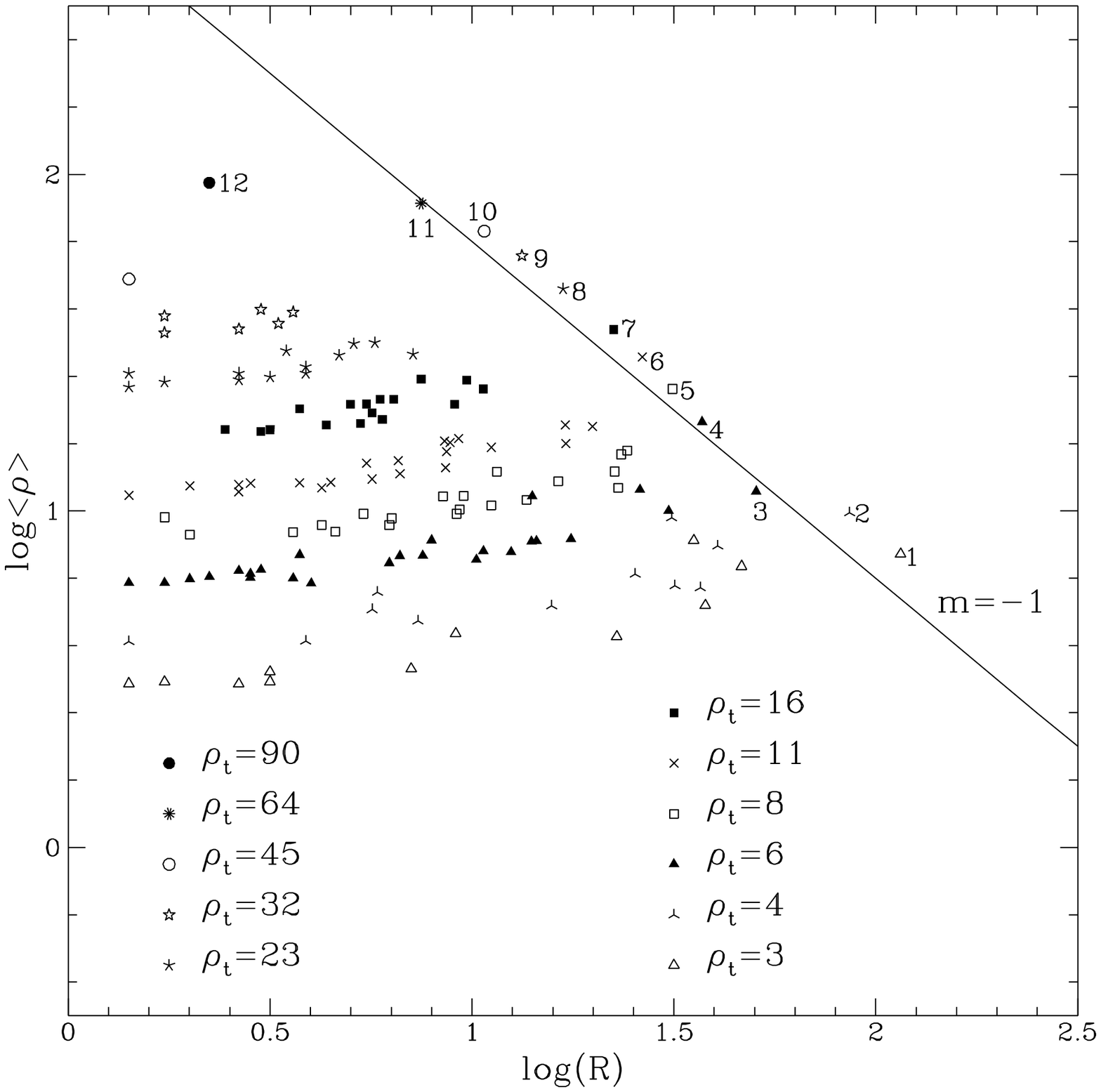}
\caption{Average cloud density {\it vs.} size for for all clouds in
a) Run 28 at $t=6.6 \times
10^7$ yr and b) Run 28bis at $t=7.15 \times 10^7$ yr. The lines show a $\rho
\propto R^{-1}$ power law. In both cases, the largest cloud at each $\rhot$
lies near the $\rho \propto R^{-1}$ line, although other times tend to show
shallower envelopes (see fig.\ \protect\ref{large clouds}). The full ensemble
of clouds does not show any trend of $\langle \rho \rangle$
with $R$. The cloud labels in a) and b)
are respectively the same as in a) and b) of figs.\ \protect\ref{larvel plot}a
and
b.
\label{lardens plot}}
\end{figure}

\begin{figure}
\plotone{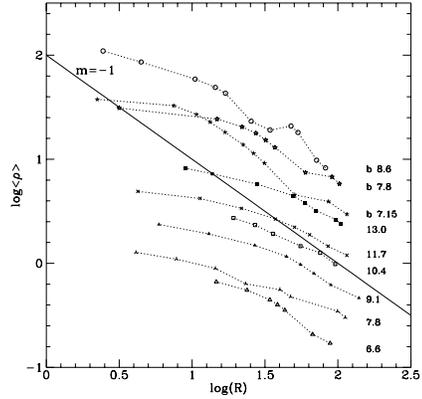}
%\vskip -5cm
\caption{Average density {\it vs.} size for
the largest cloud at each value of $\rhot$
at six different times in Run 28 (bottom curves)
and three times in Run 18bis (top curves, labeled ``b''). Each point denotes
the largest cloud at one value of $\rhot$.
Different values of $\rhot$ for the same
time in the simulation are joined by dotted lines and indicated with the same
symbol. The various curves are displaced by increments of 0.2 in $\log \rho$
for
clarity. The numbers next to each curve show the time into the evolution of the
run to which it corresponds, in units of $10^7$ yr.
Note that time $t=6.6 \times 10^7$ yr for Run 28 and time
$t=7.15 \times 10^7$ yr for Run 28bis have some of the steepest slopes. The
straight line shows a $\rho \propto R^{-1}$ power law. The average slope for
all times is $-0.81$, with a standard deviation of 0.16.
Because of the ``compression'' at high
densities due to various numerical effects, only clouds
with $\log R > 1$ (Run 28bis) or $\log R> 1.5$ (Run 28bis) were considered in
the calculation of the average slope (see text).
\label{large clouds}}
\end{figure}

\begin{figure}
\plotone{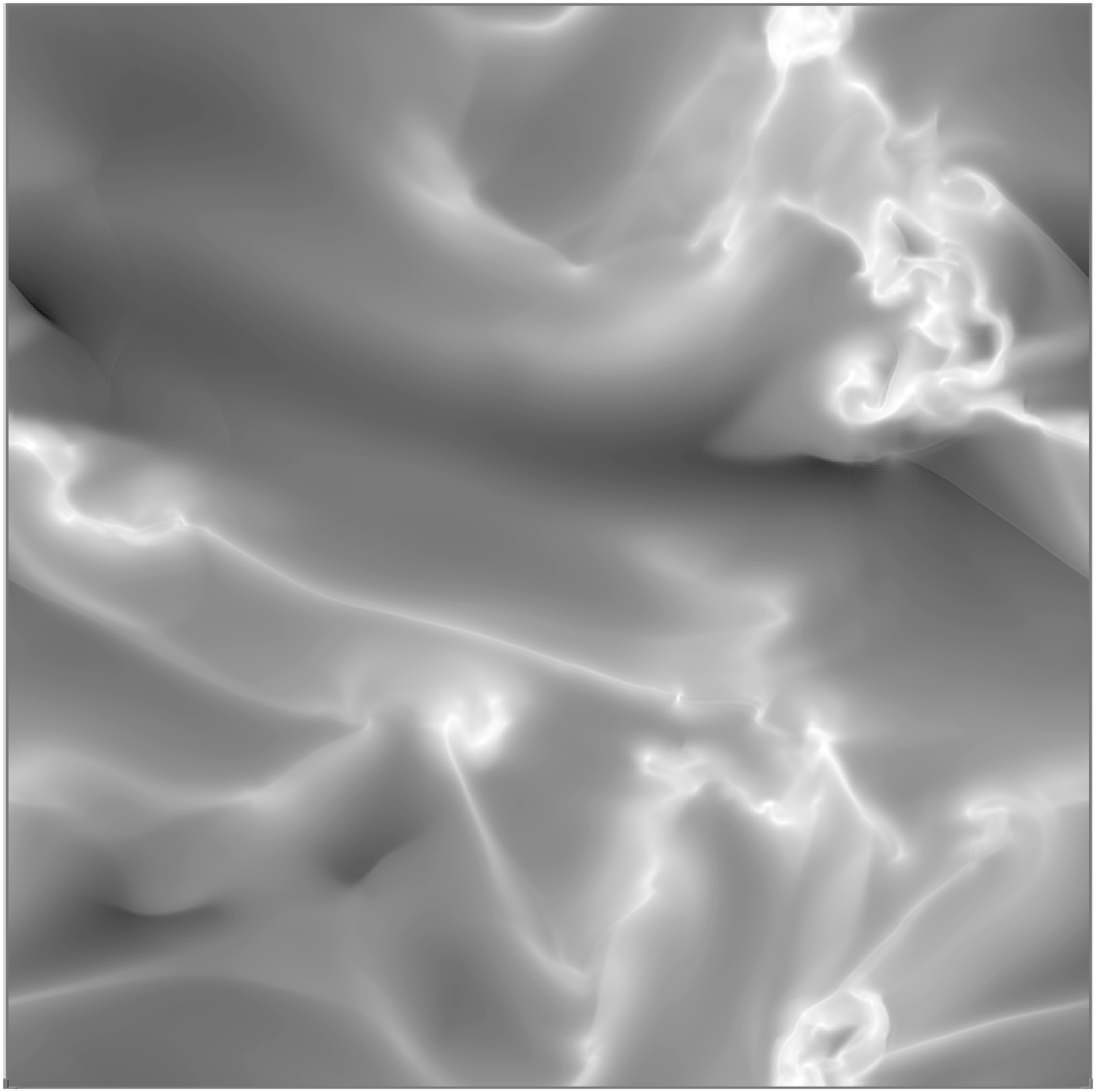}
\caption{Logarithmic grayscale image of the density field of run 28.800
  at time $t=7.15 \times 10^7$ yr. This run is similar
  to run 28bis, except that it has larger resolution ($800\times 800$), and
that
  star formation is turned off at time $t=6.9 \times 10^7$ yr. The
  density extrema are $\rho_{\rm max}=129$, and $\rho_{\rm min}=3.4
  \times 10^{-2}$.
\label{dens28.800}}
\end{figure}

\begin{figure}
\plotone{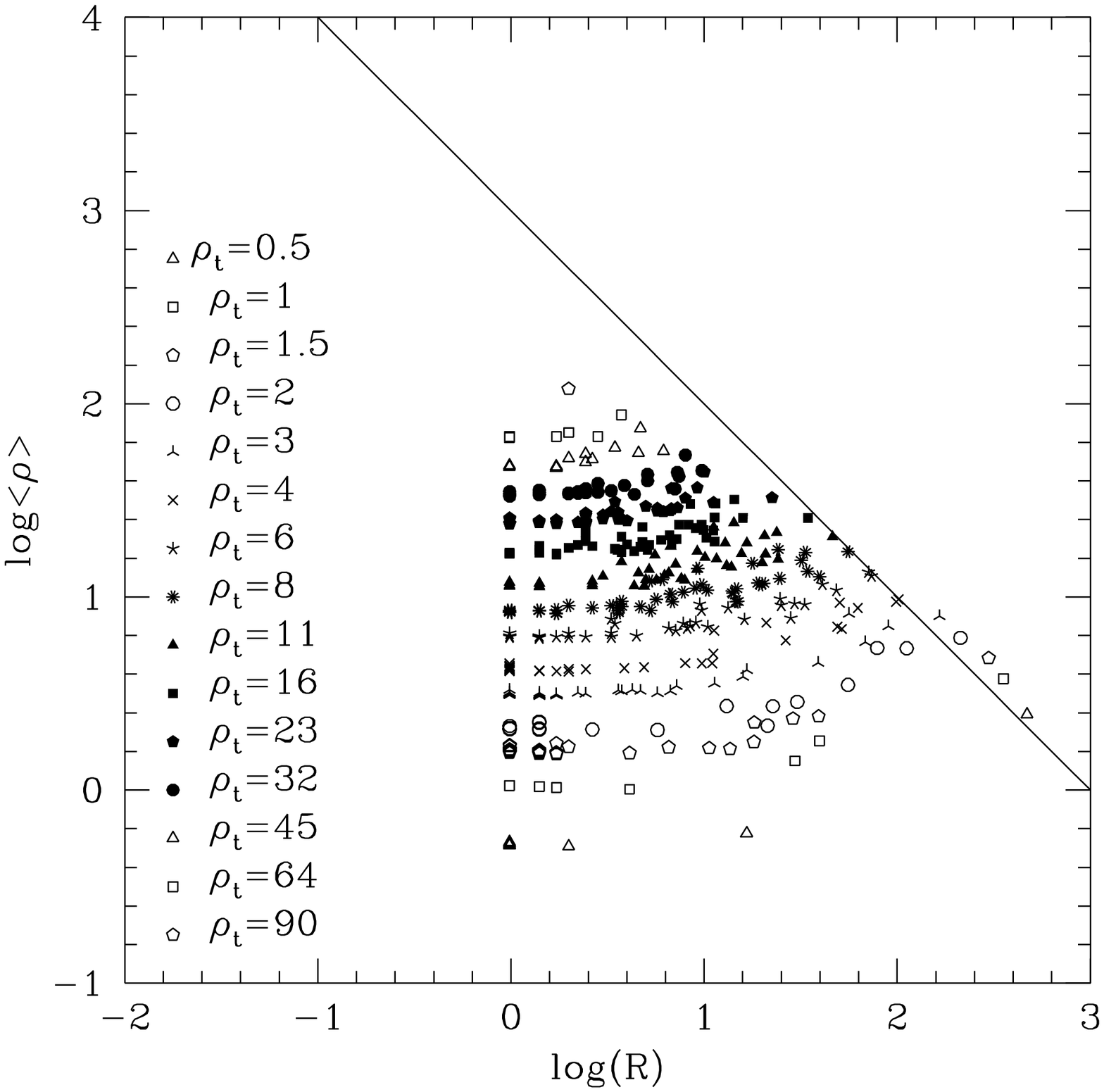}
\caption{Mean density-size plot, equivalent to figs.
  \protect \ref{lardens plot}a
  and b, but for the high-resolution Run 28.800. A large range of
  sizes, in this case spanning over two and a half orders of
  magnitude, is again seen at the lowest values of the average cloud density.
\label{lardens800 raw}}
\end{figure}

\begin{figure}
\plotone{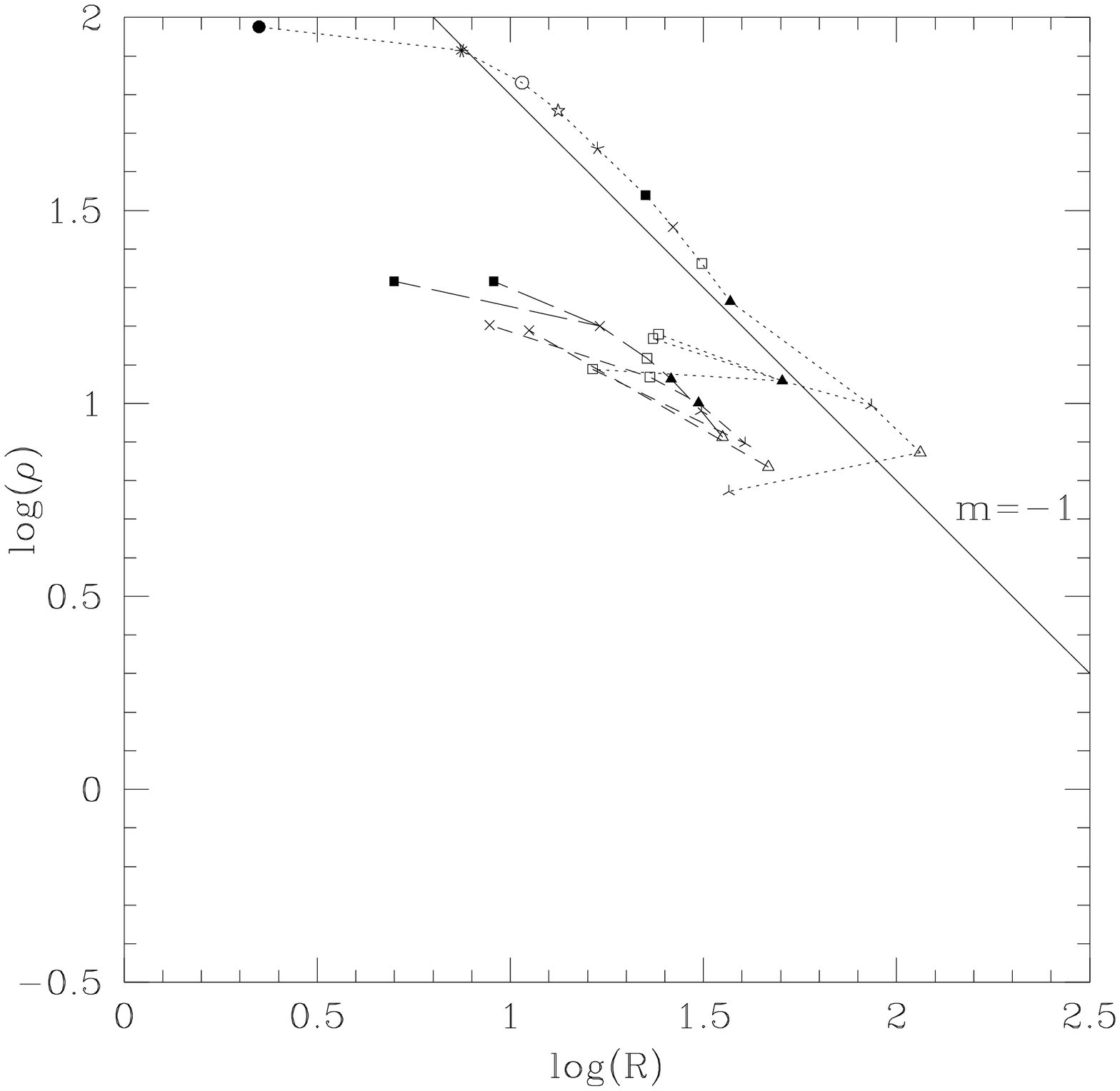}
\caption{``Familiy tree'' of a few selected clouds from fig. \protect
\ref{lardens plot}b. The largest cloud is seen to branch off to both
clouds defining a Larson-type relation as well as to clouds away from
it (dotted lines). A similar branching pattern is observed for clouds
of lower column densities (dashed lines).
\label{hierstr}}
\end{figure}

\begin{figure}
\plotone{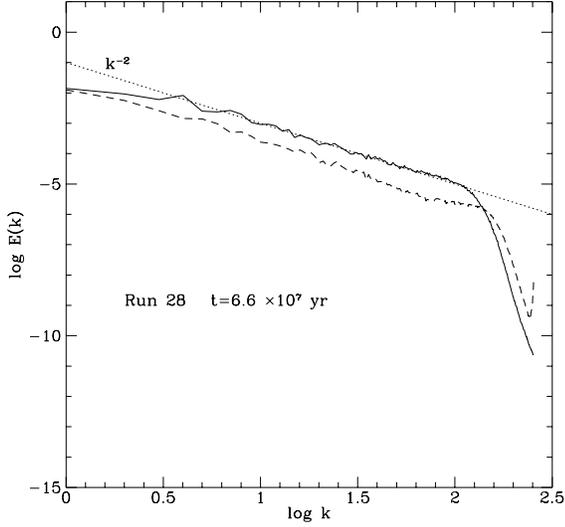}
\caption{Turbulent energy spectra for Run 28 at $t=6.6 \times 10^7$ yr.
{\it Solid} line: incompressible component of the spectrum.
{\it Dashed} line: compressible component. The straight line shows
a $k^{-2}$ power law, characteristic of shocks, which implies an $R^{-1/2}$
dependence of the velocity dispersion.
\label{spectra}}
\end{figure}

\begin{figure}
\plotone{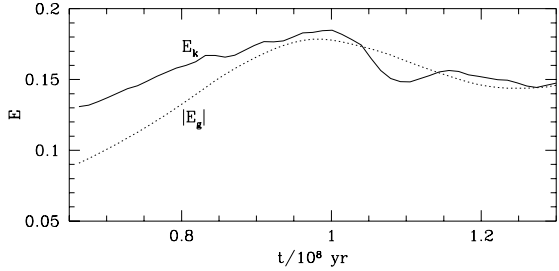}
\caption{Evolution of the total turbulent kinetic ({\it solid line})
  and the gravitational ({\it dotted line}) energies per unit mass in
  code units for run 28 over the last half of its evolution.
  The two energies are very close to equipartition at all times.}
\label{egravkin}
\end{figure}

\begin{figure}
\plotone{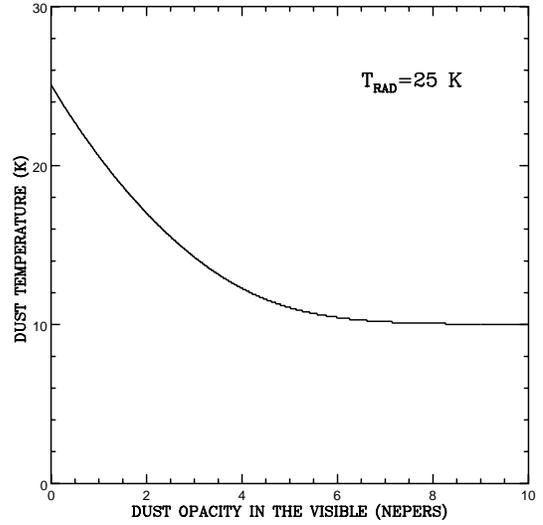}
\caption{Dust temperature, T$_{\rm DUST}$,
as a function of dust opacity in the visible
for T$_{\rm RAD}$ = 25 K. Near the edge of the cloud (left side)
T$_{\rm DUST}$ tends to T$_{\rm RAD}$, while toward the inner regions of the
cloud
(right side) heating by radiation becomes negligible
and T$_{\rm DUST}$ tends to the value provided by cosmic ray heating (10 K
in our model).
\label{figluis1}}
\end{figure}

\begin{figure}
\plotone{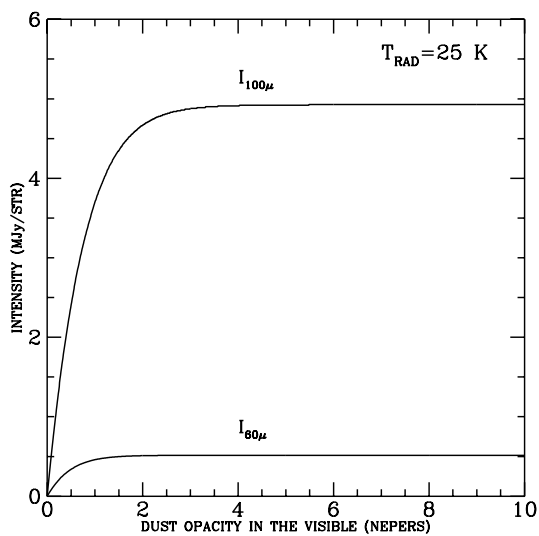}
\caption{Intensity of cloud emission at 100~$\mu$m and 60~$\mu$m as a
function of dust opacity in the visible, for
T$_{\rm RAD}$=25 K. Note that the growth of the intensity ``saturates''
above a few Nepers of dust opacity in the visible. The reason for
this effect is that beyond a few Nepers of dust opacity in the visible
there is no significant radiation heating and the dust
becomes too cold to emit significantly at 100~$\mu$m and 60~$\mu$m.
\label{figluis2}}
\end{figure}

\begin{figure}
\plotone{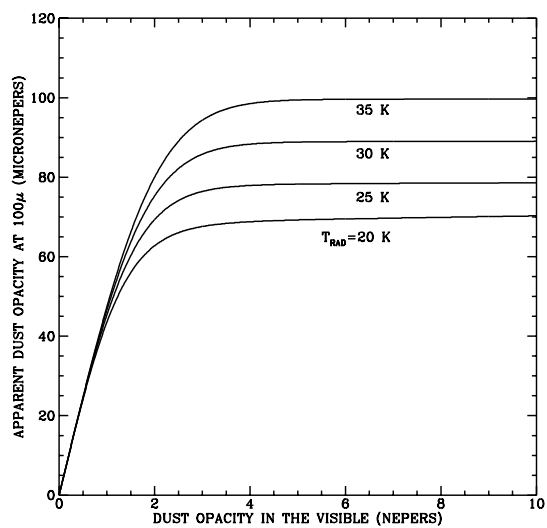}
\caption{Apparent warm dust opacity at 100~$\mu$m (derived
from the intensities at 100~$\mu$m and 60~$\mu$m),
as a function of dust opacity in the visible for
different values of T$_{\rm RAD}$. Note the ``constancy'' in
the apparent dust opacity at 100~$\mu$m once the cloud
exceeds a few Nepers in dust opacity in the visible,
practically independently of T$_{\rm RAD}$.
\label{figluis3}}
\end{figure}

\begin{figure}
\plotone{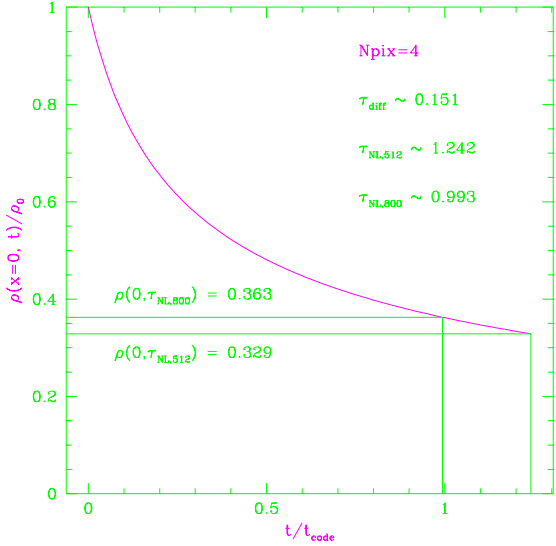}
\caption{Fractional change of the central density
$\rho (0,t)/\rho _0$ of a cloud of size 4 pixels.
The fractional change shown here is an upper bound
to the corrections applied to the simulated data,
since we have considered a cloud with size smaller
than the minimum cloud size retained, and $m=1$ in
eq. (\protect \ref{limite}).
\label{denstime}}
\end{figure}

\begin{figure}
\plotone{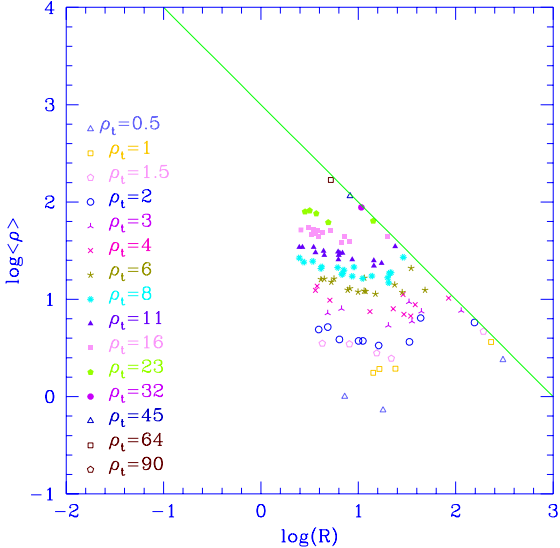}
\caption{``Corrected" density-size plot for
Run 28bis using the estimates for the `true" density
and size of the clouds given by equations
(\protect\ref{density_correction}) and (\protect
\ref{size_correction}). Note that a range of
roughly two orders of magnitude in size remains for the lowest
mean density clouds.
\label{correction}}
\end{figure}

\begin{figure}
\plotone{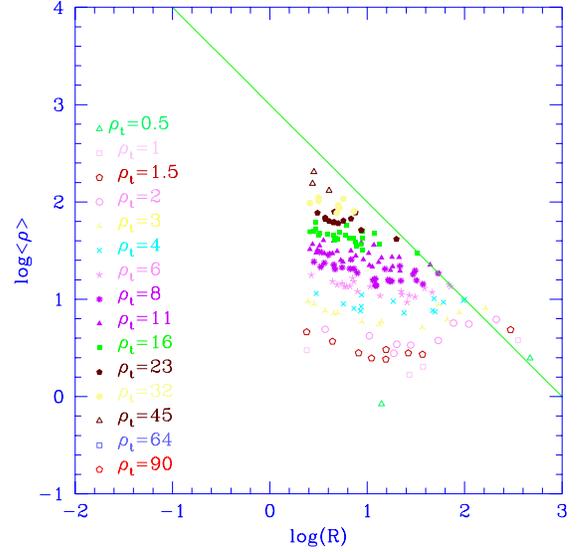}
\caption{Same as figure \protect \ref{correction}, but for a
Run~28.800. Here, the sizes of the smallest-density clouds vary by nearly 2.5
orders of magnitude. Note that, although only clouds with sizes larger
than 4 pixels are retained, the smallest clouds in this figure have
sizes smaller than that because of the size ``correction'', eq.
(\protect \ref{size_correction}).
\label{correction_800}}
\end{figure}
\end{document}